\documentclass[final,3p,times,twocolumn,numbers]{elsarticle}

\usepackage{amssymb}
\usepackage{amsmath}
\usepackage[ruled,vlined,linesnumbered]{algorithm2e}
\SetKwInput{KwIn}{Require}
\SetKwInput{KwOut}{Ensure}
\usepackage{caption}
\usepackage{subcaption}
\usepackage{xcolor}    
\usepackage[colorlinks=true,
            citecolor=blue,
            linkcolor=black,
            urlcolor=blue]{hyperref} 
\usepackage{fontawesome5}
\usepackage{bm}
\usepackage{enumitem}
\usepackage{float}
\usepackage{placeins}
\usepackage{titlesec}
\usepackage{caption}
\usepackage{multirow}
\usepackage{etoolbox}
\hypersetup{
    colorlinks=true,
    citecolor=blue,
    linkcolor=black,
    urlcolor=blue
}
\usepackage[numbers]{natbib}
\titleformat{\paragraph}[runin]
{\itshape}
{}
{0pt}
{}

\begin{document}

\begin{frontmatter}



\title{Optimizing a discrete loss with finite-difference-based physical constraints and time-stepping for solving PDEs.}


\author[1,2,4]{Yali Luo}
\author[1,4]{Yiye Zou}
\author[5,6]{Heng Zhang}
\author[4]{Mingjie Zhang}
\author[5]{Gang Wei}
\author[3,4,5]{Jingyu Wang\corref{cor1}} 
\author[4]{Xiaogang Deng}
\cortext[cor1]{Corresponding author.
\newline E-mail address: wangjingyu@scu.edu.cn(Jingyu Wang)}  
\affiliation[1]{{School of Computer Science, Sichuan University,}, city={Chengdu 610065}, country={China}}
\affiliation[2]{{Chengdu Jincheng College,},city={Chengdu 611731}, country={China}}
\affiliation[3]{{School of Aeronautics and Astronautics,Sichuan University,},city={Chengdu 610065},country={China}}
\affiliation[4]{{National Key Laboratory of Fundamental Algorithms and Models for Engineering Simulation, Sichuan University,},city={Chengdu 610207},country={China}}
\affiliation[5]{{Simulation and Precision Control Research Center, Taihang Laboratory,},city={Chengdu 610213}, country={China}}
\affiliation[6]{{Institute for Aero Engine, Tsinghua University,},city={Beijing 100084}, country={China}}

\begin{abstract}
Computational Fluid Dynamics (CFD) is an important approach for analyzing flow phenomena and predicting engineering-relevant quantities. The governing physics is typically formulated as partial differential equations (PDEs) and solved numerically on computational grids. Physics-informed
    neural networks (PINNs) have emerged as a popular optimization-based approach for solving PDEs, but they often suffer from ill-conditioned objectives and the high cost of automatic differentiation. Optimization-based discretizations (ODIL) mitigate several PINN drawbacks by optimizing discrete variables directly, yet accuracy and efficiency remain limited on body-fitted geometries and for time-dependent problems. This paper proposes a finite-difference-based time-stepping loss optimization solver (FDTO) that defines physics losses from discrete residuals. FDTO introduces a novel approach by applying curvilinear coordinate transforms with body-fitted structured grids and decomposes long-horizon evolution into sequential, well-conditioned subproblems consistent with time marching. The method is primarily evaluated on incompressible Navier--Stokes flows, including lid-driven cavity
    benchmarks, external airfoil aerodynamics (lift/drag consistency), and a cylinder case on a multi-block structured grid with cross-block coherent solutions. Additional validations on diffusion and flow-mixing problem further demonstrate generality. Compared with representative PINN-based solvers, FDTO reduces GPU memory by about
    82.6\% on the lid-driven cavity case and achieves 3--5 times lower relative error on the flow-mixing problem. These results indicate that FDTO enables accurate, stable, and memory-efficient discrete-loss optimization for incompressible-flow solutions, while remaining applicable to other PDE models. The implementation code of this paper is available on GitHub at
    \url{https://github.com/My-git96/FDTO\_PDE}.

\end{abstract}




\end{frontmatter}




\section{Introduction}
\label{introduction}
In scientific computing and engineering, computational fluid dynamics (CFD) studies flow phenomena by formulating the governing physics as partial differential equations (PDEs). These PDEs are then solved numerically by discretizing them on computational grids, yielding algebraic systems whose solutions approximate the underlying physical behavior~\citep{blazek2015computational}. After decades of development, numerical solution methods such as the finite-difference method (FDM)~\citep{leveque2007finite}, finite-volume method (FVM)~\citep{moukalled2015finite}, and finite-element method (FEM)~\citep{reddy1993introduction},have achieved tremendous success~\citep{wang2018high,wang2020blending,chen2024separation}. However, traditional numerical methods often exhibit limited computational efficiency in the presence of complex geometries and boundary conditions~\citep{RQWL202502007,RQWL202504014}.

Advances in computing hardware and data science have promoted data-driven approaches for physics-based problems~\citep{maruyama2021data,huang2024neural,xu2023practical}. Deep neural networks provide new opportunities for efficient PDE solvers~\citep{lu2026amspinn}. Recent studies have explored integrating neural networks with PDE methods to enhance computational performance and to learn spatiotemporal dynamics from data~\citep{bar2019learning,sanchez2020learning,thuerey2020deep}. Yet the increasing prevalence of noisy or incomplete data, together with the high cost of high-quality datasets, calls for more flexible and robust solution strategies~\citep{chen2024towards}.

In recent years, optimization-based PDE solvers that minimize scalar objectives have attracted substantial attention. In this paradigm, unknown fields can be parameterized by neural networks and optimized by minimizing residual-based objectives, as in physics-informed neural networks (PINNs)~\citep{raissi2019physics}, or treated as discrete degrees of freedom and optimized directly through discrete residual loss, as in discrete-loss optimization (ODIL)~\citep{karnakov2024solving}.

PINNs provide a grid-free paradigm for solving PDEs and can be trained without labeled solution data by enforcing governing equations and boundary/initial conditions through a physics-informed objective. With automatic differentiation (AD)~\citep{baydin2018automatic}, PINNs evaluate PDE residuals and optimize network parameters to satisfy the governing equations and boundary conditions. This physics-constrained learning paradigm has shown promise across a range of PDEs, including the Navier--Stokes equations~\citep{rao2020physics}. To alleviate the ill-conditioned global objectives in standard PINNs, the time-stepping oriented neural network (TSONN) reformulates the optimization in a time-marching manner, embedding classical time-stepping schemes into the learning process~\citep{cao2025analysis}. Nevertheless, PINNs typically rely on AD in deep-learning frameworks to compute derivatives of the physical fields. When high-order derivatives are required, AD can become prohibitively expensive during backpropagation because it must generate and retain a large number of intermediate tensors~\citep{bhatia2024lowering,gunes2015automatic}. As a result, PINNs are often difficult to scale to PDEs involving high-order differential operators, since the cost of nested AD grows rapidly with the differentiation order~\citep{baydin2018automatic,bettencourt2019taylor}. Moreover, for complex flows at high Reynolds numbers, PINNs may converge to nonphysical pseudo-solutions~\citep{wang2023solution}. 

These limitations have motivated new hybrid approaches that integrate deep learning with classical numerical discretizations, using established schemes to approximate derivatives while retaining the flexibility of learning-based representations~\citep{li2024predicting,lu2025unsupervised,gao2022physics}. Some works propose a hybrid approach incorporating
stabilized FEM with PINN to leverage the
robustness of classical discretizations ~\citep{cengizci2026pinn}.
Others extend the PINN-Augmented SUPG with
Shock-Capturing (PASSC) methodology from steady to unsteady problems, combining a semi-discrete stabilized FEM with a PINN-based correction strategy for transient convection-diffusion-reaction equations ~\citep{cengizci2026physics}.
While these hybrid methods show promise, training neural networks can be computationally intensive. In addition, when neural networks are embedded within PDE solvers, convergence must be monitored not only through reductions in PDE residuals but also through the optimization of network parameters. This coupling complicates convergence diagnostics and may obscure whether the computed solution has truly converged.

A complementary optimization-based approach is ODIL, which builds on conventional PDE discretizations and does not rely solely on a neural surrogate to represent the solution. Instead, it formulates the governing equations on discretized variables and minimizes a loss constructed from the discretized PDE residuals. The resulting optimization problem can be solved using gradient-based or Newton-type methods. By inheriting the structure and properties of the underlying discretization, ODIL better preserves physical consistency and numerical stability, mitigating ill-conditioned global objectives, unstable gradients, and AD-related overhead that commonly hinder PINNs. In addition, ODIL can be combined with modern machine-learning techniques to accelerate selected numerical components and improve overall computational efficiency. ODIL has been successfully applied to a range of incompressible viscous-flow settings~\citep{karnakov2023flow}. More recently, the ODIL framework has been extended to jointly infer both obstacle geometry and the associated flow field in three-dimensional steady-state supersonic flows~\citep{buhendwa2025data}. In parallel, efforts to improve ODIL efficiency have focused on alleviating loss ill-conditioning: Cao and Zhang~\citep{cao2025overcoming} proposed the stabilized gradient residual (SGR) loss, which is better conditioned than mean-squared residual objectives and can substantially accelerate convergence.

While ODIL has shown strong performance as an optimization-based PDE solver, existing ODIL formulations have been developed and demonstrated primarily on uniform Cartesian grids, to the best of current knowledge~\citep{karnakov2024solving,cao2025overcoming}. In practical engineering applications, especially for time-dependent flows with pronounced boundary-layer effects, body-fitted grid with near-wall refinement are typically essential. Extending discrete optimization-based solvers to such body-fitted geometries therefore remains an important challenge. Beyond geometric considerations, many physical systems evolve in both space and time.
In fluid mechanics, these dynamics are governed by the Navier--Stokes equations ~\citep{temam2024navier}, requiring solvers to accurately capture temporal evolution in addition to spatial structure. Most existing ODIL studies, however, have focused on computing time-invariant solutions or single-time objectives~\citep{karnakov2024solving, cao2025overcoming}, which limits applicability to convection-dominated flows where the solution undergoes sustained temporal development and complex wake-boundary-layer interactions. These regimes typically demand stable time marching, motivating an ODIL framework that performs trajectory-level optimization over multiple time steps rather than optimizing a single stationary residual.
Furthermore, prior ODIL results have been reported mainly on canonical benchmarks such as the lid-driven cavity, where efficiency is typically demonstrated on simple geometries using a single structured grid. The behavior of ODIL on body-fitted multi-block grids and in more realistic external-flow configurations remains less explored.

Therefore, this paper presents a finite-difference-based time-stepping loss optimization solver (FDTO) that addresses the challenges of geometric adaptability and efficient time-dependent flow. By incorporating scheme-consistent finite-difference operators in curvilinear coordinates, this work develops a discrete-loss optimization for body-fitted multi-block structured grids. Furthermore, a time-stepping-oriented optimization strategy is proposed to improve conditioning, stability, and efficiency, while maintaining physical consistency. The main contributions are as follows:

\begin{itemize}
  \item \textbf{Geometrically adaptable optimization on body-fitted grids.} This paper proposes a finite-difference discrete-loss optimization scheme in curvilinear coordinates for body-fitted structured grids, including multi-block configurations. By minimizing discrete residuals with interface-consistent transfers, the resulting formulation enhances geometric and boundary-condition compatibility, while maintaining the efficiency and accuracy of finite-difference discretizations.

  \item \textbf{Time-marching optimization at the trajectory level.}
  Time-marching optimization is applied by reformulating the global time-dependent objective into a sequence of time-window subproblems consistent with the time-marching updates. This trajectory level decomposition improves conditioning and promotes stable convergence for nonlinear time-dependent optimization.

  \item \textbf{Validation on engineering-relevant configurations.}
  The proposed method is evaluated on a range of incompressible Navier--Stokes cases, including the lid-driven cavity, airfoils, and flow around a cylinder on body-fitted grids, as well as diffusion and flow-mixing problem.
  The results demonstrate strong accuracy, robustness, and memory efficiency across geometry-adaptive configurations and long-term temporal evolution. Additionally, a detailed analysis of aerodynamic parameters, such as lift, drag, and pressure distribution is also included in the study.
\end{itemize}

The remainder of this paper is organized as follows. Section~2 presents the proposed method, including the coordinate-transformed finite-difference discretization and the time-stepping optimization strategy. Section~3 reports results on incompressible Navier--Stokes equation, diffusion equation, flow-mixing problem, and compares proposed method with representative baselines. Finally, Section~4 concludes the paper and discusses limitations and future directions.
\section{Methodology}
\titleformat{\subsection}
  {\bfseries}              
  {\textbf{\thesubsection.}} 
  {0.8em}                  
  {}
\subsection{\textbf{Paradigms of PDE Solvers}}
\begin{figure*}[!ht]
\centering
\includegraphics[width=0.98\textwidth]{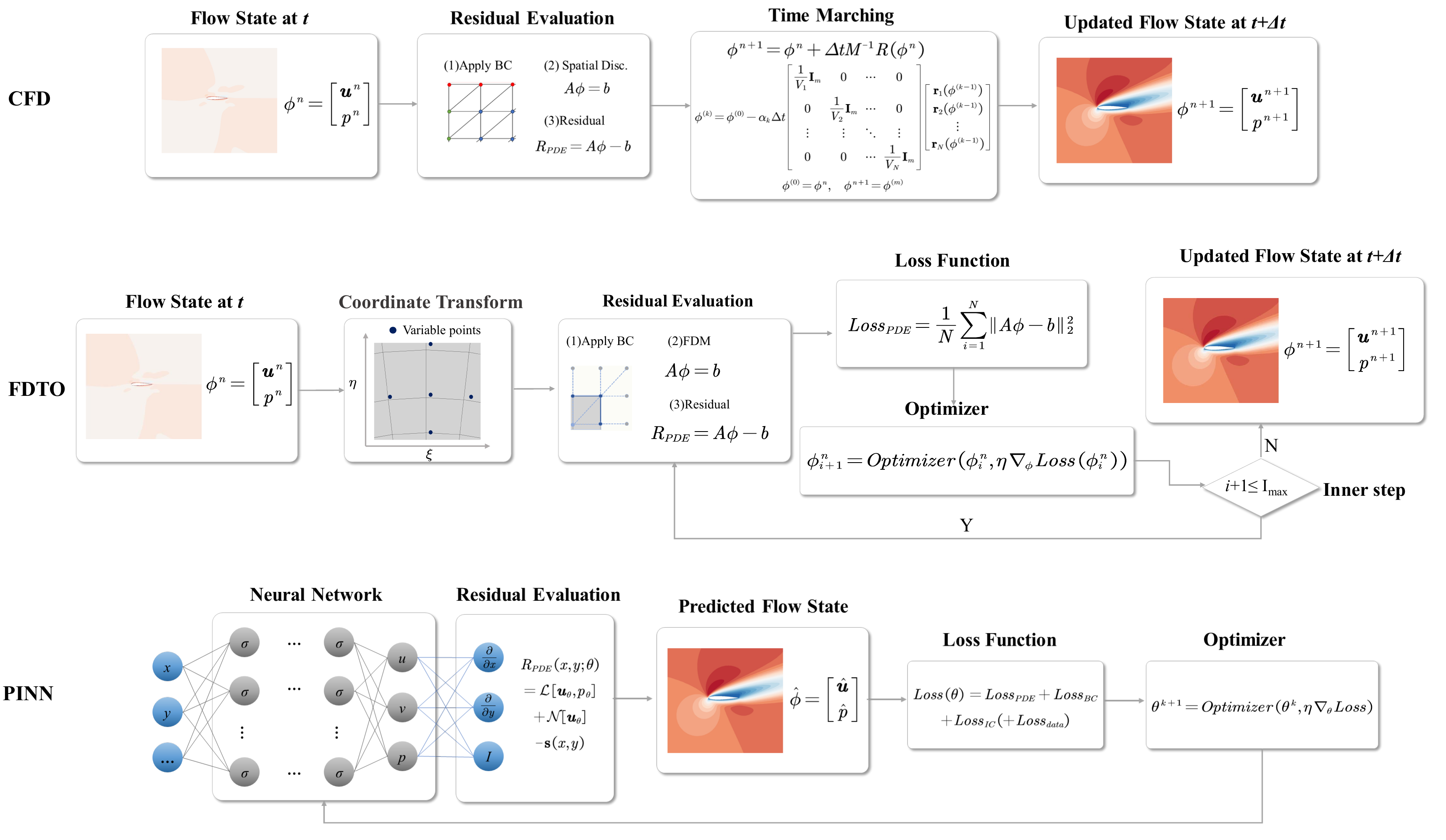}
\captionsetup{labelfont={color=blue}, textfont={color=blue}}
\caption{Paradigms of PDE solvers.
top: CFD; middle: FDTO; bottom: PINN.}
\label{fig:paradigms}
\end{figure*}
Three representative paradigms for solving PDEs in fluid dynamics are illustrated in Fig.~\ref{fig:paradigms}: conventional CFD, FDTO, and PINN. Although all three target the same governing equations, they differ fundamentally in how PDE constraints are enforced and how the solutions are represented and updated.

In conventional CFD solvers, the governing equations are discretized in space and time, and the flow field is advanced through explicit or implicit time-marching schemes. At each time step, large sparse linear or nonlinear systems must be solved, typically involving global matrix assembly and iterative solvers. While this paradigm is well established and numerically robust, its computational cost and memory footprint grow rapidly for fine grids, complex geometries, and time-dependent simulations, particularly when repeated solves are required in optimization, inverse problems, or parametric studies.

PINN reformulates PDE solving as a global function-approximation problem, where a neural network is trained to satisfy the governing equations together with boundary conditions, and initial conditions through loss minimization. This grid-free paradigm avoids explicit matrix assembly and offers strong flexibility for irregular domains. Nevertheless, PINN introduces new challenges, including high training cost, substantial memory overhead from automatic differentiation, and stiff objectives arising from competing loss terms. These issues become especially pronounced for time-dependent, convection-dominated regimes at high Reynolds numbers, where reliable convergence and accuracy are difficult to guarantee.

The proposed FDTO framework represents a unique intermediate paradigm between classical CFD and PINN. 
Similar to CFD, FDTO retains finite-difference discretizations on  body-fitted structured grids, thereby preserving the numerical stability, locality, and physical consistency of traditional schemes. In contrast to CFD, FDTO avoids the solution of large algebraic systems. Instead, the discrete flow variables are updated by minimizing a residual-based loss function at each time step, enabling matrix-free updates driven by optimization algorithms.

Compared with PINN, FDTO optimizes discrete flow variables directly rather than neural-network parameters.
This removes the need for network training, reduces the memory overhead of deep automatic-differentiation graphs, and avoids the stiffness associated with global function approximation. 
By constructing the objective from classical finite-difference operators, FDTO inherits the stability and consistency of established numerical schemes. Moreover, this discrete-optimization viewpoint supports time-marching-oriented optimization for time-dependent problems, while remaining compatible with complex geometries and conventional CFD data structures.

In summary, CFD solves the discretized governing equations, PINN learns a global neural approximation, and FDTO optimizes discrete flow states. By incorporating coordinate transformations and operates in curvilinear coordinates on body-fitted grids, FDTO improves geometric adaptability for complex geometries. Coupled with finite-difference discretization with time-stepping-oriented optimization, FDTO further achieves a favorable balance among numerical stability, computational efficiency, and modeling flexibility for time-dependent problems.
\subsection{\textbf{FDTO Framework}}
\label{sec:FDTO_framework}
The general formulation of the proposed FDTO framework is shown in Fig.~\ref{fig:paradigms}. Built on body-fitted structured grids, FDTO casts each time-marching update as a discrete optimization problem, leading to improved conditioning and enhanced numerical stability.

\textit{Body-fitted structured grid and node-wise representation:}
The physical domain is discretized with a body-fitted structured grid that conforms to the boundary geometry. Node-wise quantities are stored at each grid node, including the flow state $\phi=(\bm{u},p)$, precomputed geometric metrics (e.g., $\xi_x,\xi_y,\eta_x,\eta_y,J^{-1}$), and PDE-related parameters. This node-wise representation retains the locality and stencil structure of classical finite-difference schemes, thereby providing a natural basis for discrete optimization over the flow variables.

\begin{figure*}[!ht]
  \centering
  \includegraphics[width=0.7\linewidth]{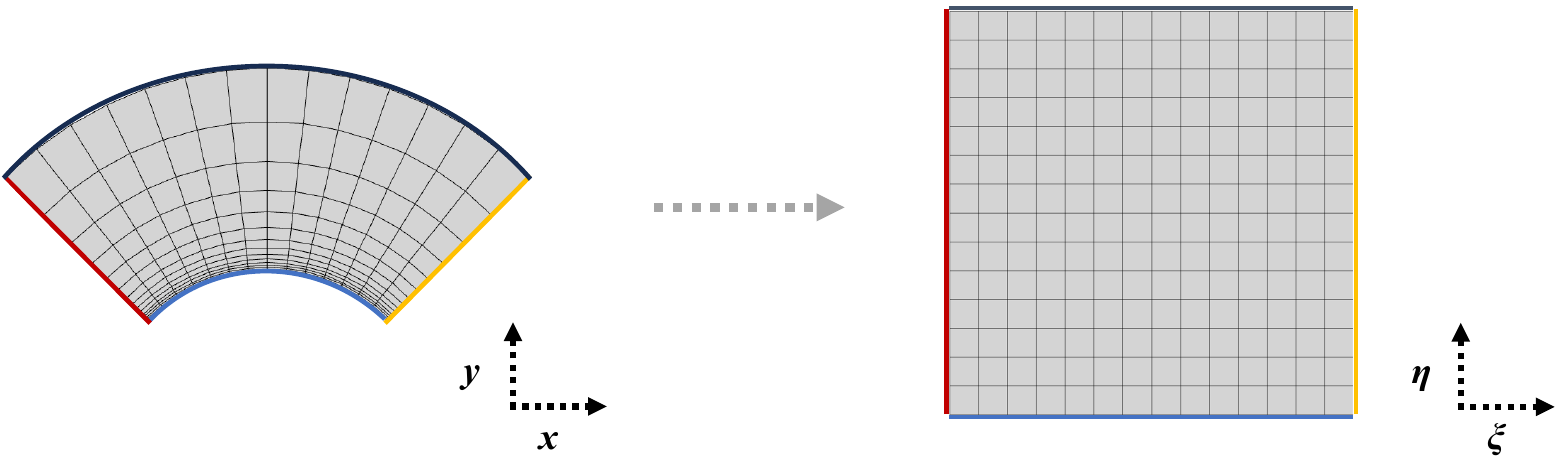}
  \caption{An example of coordinate transformation from physical space (left) to computational space (right).}
  \label{fig:coord_transform}
\end{figure*}

\textit{Curvilinear coordinate transformation:}
In two dimensions, a coordinate mapping is introduced from the physical space $(x,y)$ to the computational space $(\xi,\eta)$:
\begin{equation}
\label{eq:map_213}
\left\{
\begin{aligned}
\xi&=\xi(x,y),\\
\eta&=\eta(x,y).
\end{aligned}
\right.
\end{equation}
Fig.~\ref{fig:coord_transform} illustrates a coordinate mapping from the physical domain to the computational domain, where a body-fitted curvilinear grid in the physical space is mapped onto a uniform grid in the computational space. Curves in the same color denote corresponding grid lines (or boundary curves) in the two coordinate systems. Throughout, subscripts denote partial derivatives, for example, $q_x=\partial q/\partial x$.
The Jacobian matrix associated with the mapping can be written as
\begin{equation}
\label{eq:jac_214}
J=
\begin{bmatrix}
\xi_x & \eta_x\\
\xi_y & \eta_y
\end{bmatrix},
\end{equation}
and its inverse is
\begin{equation}
\label{eq:jac_inv_214}
J^{-1}
=
\begin{bmatrix}
x_\xi & y_\xi\\
x_\eta & y_\eta
\end{bmatrix}.
\end{equation}
where the determinant of the inverse Jacobian matrix is
\begin{equation}
\label{eq:jac_det}
\det(J^{-1})=x_\xi y_\eta-x_\eta y_\xi.
\end{equation}
Accordingly, the metric coefficients (grid derivatives) satisfy:
\begin{equation}
\left\{
\begin{aligned}
\tilde{\xi_x} &= {\xi_x}/{\det(J^{-1})} = y_\eta,\\
\tilde{\xi_y} &= {\xi_y}/{\det(J^{-1})} = -x_\eta,\\
\tilde{\eta_x} &= {\eta_x}/{\det(J^{-1})} = -y_\xi,\\
\tilde{\eta_y} &= {\eta_y}/{\det(J^{-1})} = x_\xi.
\end{aligned}
\right.
\end{equation}
In practice, the geometric derivatives $x_\xi,x_\eta,y_\xi$ and $y_\eta$ are evaluated on the structured grid using finite differences. In this work, these partial-derivative terms are approximated using second-order central finite differences at interior points, while second-order forward and backward finite differences are used near boundaries.
The computational grid spacings are often taken as $\Delta \xi=\Delta \eta=1$.
The coordinate transformation of the PDEs is formulated in a conservative form in computational space. A general two-dimensional conservation-law PDE is considered in the physical space $(x,y)$.
\begin{equation}
\label{eq:cons_phys_216}
\frac{\partial Q}{\partial t}
+
\frac{\partial F}{\partial x}
+
\frac{\partial G}{\partial y}
=0,
\end{equation}
where $Q$ is the conserved quantity, and $F$ and $G$ are the fluxes in the $x$- and $y$-directions, respectively.
Given the coordinate mapping in Eq.~\eqref{eq:map_213}, physical-space derivatives can be written via the chain rule as 
\begin{equation}
\label{eq:chain_217}
\frac{\partial}{\partial x}
=
\xi_x\frac{\partial}{\partial \xi}
+
\eta_x\frac{\partial}{\partial \eta},
\qquad
\frac{\partial}{\partial y}
=
\xi_y\frac{\partial}{\partial \xi}
+
\eta_y\frac{\partial}{\partial \eta}.
\end{equation}
Applying Eq.~\eqref{eq:chain_217} to Eq.~\eqref{eq:cons_phys_216} results in
\begin{equation}
\label{eq:cons_comp_218}
\frac{\partial Q}{\partial t}
+
\frac{\partial}{\partial \xi}\!\left(\xi_x F+\xi_y G\right)
+
\frac{\partial}{\partial \eta}\!\left(\eta_x F+\eta_y G\right)
=0.
\end{equation}
To retain a conservative form in computational space, the equation is written as
\begin{equation}
\label{eq:cons_final_219}
\frac{\partial\!\left(J^{-1}Q\right)}{\partial t}
+
\frac{\partial \tilde{F}}{\partial \xi}
+
\frac{\partial \tilde{G}}{\partial \eta}
=0,
\end{equation}
where the transformed (contravariant) fluxes are defined by
\begin{equation}
\label{eq:flux_def_220}
\tilde{F}=\tilde{\xi_x}F+\tilde{\xi_y}G,
\qquad
\tilde{G}=\tilde{\eta_x}F+\tilde{\eta_y}G.
\end{equation}
Eq.~\eqref{eq:cons_final_219} is thus the conservative-form representation of the conservation law in the curvilinear computational coordinates $(\xi,\eta)$.

\textit{Boundary Condition:} After the coordinate transformation, appropriate boundary closures are applied to ensure a well-posed discrete formulation on body-fitted structured grids. FDTO employs explicit ghost nodes, and boundary treatment is constructed locally using three node sets (Fig.~\ref{fig:paradigms}): ghost nodes $\mathcal{G}$ outside the physical domain, boundary nodes $\mathcal{B}$ on the physical boundary, and adjacent interior nodes $\mathcal{I}$ involved in near-boundary stencils. For each boundary segment, ghost-node states are populated by a linear extrapolation operator that uses information on $\mathcal{B}$ and $\mathcal{I}$ to maintain stencil consistency for gradient evaluation near the boundary. Specifically, for a generic variable $\phi\in\{\bm{u},p\}$,
\begin{equation}
\label{eq:ghost_extrap}
\phi_{\mathrm{ghost}}=\mathcal{E}_{\mathrm{lin}}\!\left(\phi_{\mathcal{B}},\,\phi_{\mathcal{I}}\right),
\end{equation}
where $\mathcal{E}_{\mathrm{lin}}(\cdot)$ denotes a linear extrapolation along the local boundary-normal direction (equivalently, along the boundary-adjacent grid line). As illustrated by the dashed diagonal in Fig.~\ref{fig:paradigms}, each boundary node is paired with its nearest interior neighbor along this local normal direction, and the ghost-node value is obtained by linearly extending the same line beyond the boundary. This construction yields a smooth, geometry-consistent closure for near-boundary stencils and provides a unified discrete treatment across different physical boundary specifications without introducing additional penalty terms. Consequently, ghost-node extrapolation preserves locality, maintains consistent residual evaluation near boundaries, and improves robustness during time-stepping optimization.

\begin{figure*}[!ht]
\centering
  \includegraphics[width=0.7\linewidth]{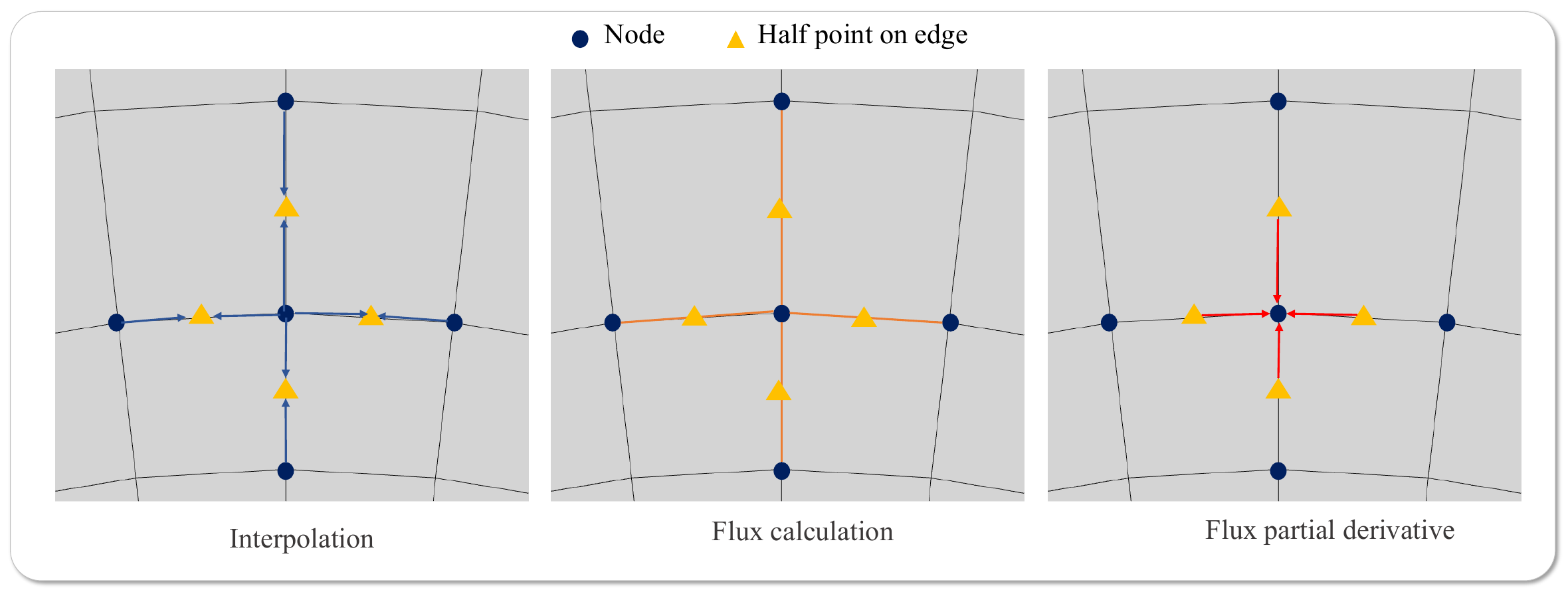}
  \caption{Diagram of FDTO spatial discretization procedure. (left) Node-stored variables (and precomputed grid metrics) are linearly interpolated to edge midpoints to obtain face states. (middle) Conservative numerical fluxes are evaluated on these faces using the interpolated states. (right) The nodal residual is assembled by a finite-difference flux divergence, i.e., differences of neighboring face fluxes in the computational $\xi$/$\eta$ directions.}
  \label{fig:fdm}
\end{figure*}

\textit{Finite-difference discretization:} Spatial discretization is performed using conservative finite-difference schemes. As illustrated in Fig.~\ref{fig:fdm}, interpolation, flux evaluation, and flux divergence are applied at distinct stencil locations, leading to discrete residuals that closely mirror those of classical CFD solvers. Boundary conditions are imposed directly at the discrete level by modifying the residuals associated with boundary nodes, preserving physical constraints without introducing additional penalty terms. Further implementation details of the finite-difference operators and stencil organization please refer to finite difference graph network (FDGN)~\citep{zou2024finite}.
Let $N_h$ denote the number of nodes of a spatial grid in $\mathbb{R}^d$.
The nodal unknowns are stacked as:
\begin{equation}
\phi
=
\big[
\bm{u}_0,\,p_0,\,
\bm{u}_1,\,p_1,\,
\ldots,\,
\bm{u}_{N_h-1},\,p_{N_h-1}
\big]^{\top}
\in \mathbb{R}^{3N_h}.
\label{eq:q_stack}
\end{equation}
where $\bm{u}_{j}=[u_{j},\,v_{j}]^{\top}$ and $p_{j}$ denote the velocity vector and pressure at the $j$-th node, respectively, for $j=0,\ldots,N_h-1$.
A class of nonlinear PDEs is considered in the form:
\begin{equation}
\mathcal{N}[\phi](x,y,t)=f_s(x,y,t), \qquad (x,y)\in\Omega,t\in{\Gamma},
\label{eq:pde}
\end{equation}
where $\mathcal{N}$ denotes a nonlinear differential operator, $\phi(x,y)$ is the unknown solution field, and $f_s(x,y)$ is the source term. The physical domain $\Omega$ is discretized and $\mathcal{N}$ is approximated by finite differences scheme, yielding a system of nonlinear algebraic equations that can be written compactly as a discrete residual:
\begin{equation}
\mathbf{r}_{\mathrm{PDE}}(\phi)=0. \qquad 
\label{eq:discrete_residual}
\end{equation}

\textit{Residual-based loss formulation:} A locally linearized residual system is considered, given by:
\begin{equation}
f=\mathbf{A} \phi-\mathbf{b}=0,
\end{equation}
where $\mathbf{A} \in \mathbb{R}^{n \times n}$ is the Jacobian matrix (or linearized residual operator), $\phi \in \mathbb{R}^n$ is the discretized solution vector, and $\mathbf{b}  \in \mathbb{R}^n$ is the source term evaluated at the current iterate $\phi^{n}$, with $n$ representing the time step. 

The finite-difference discretization leads to the residual system
\begin{equation}
\mathbf{r}_{\mathrm{PDE}}(\phi)=\mathbf{A}\phi-\mathbf{b},
\end{equation}
which is incorporated into the loss function
\begin{equation}
\mathcal{L}_{\mathrm{PDE}}(\phi)
=
\frac{1}{N_h}
\sum_{j=0}^{N_h-1}
\left\lVert r_{\mathrm{PDE}, j}(\phi) \right\rVert^2 .
\end{equation}
This approach avoids the need for global matrix inversion, while maintaining the numerical consistency of the underlying discretization scheme.

The FDTO minimizes the loss function using gradient-based optimizers, such as Adam~\citep{kingma2014adam}, SOAP~\citep{vyas2024soap}, or L-BFGS~\citep{zhu1997algorithm}. The full loss function is given by
\begin{equation}
\mathcal{L}_{\mathrm{PDE}}(\phi)=\frac{1}{N} \sum_{i=1}^{N} \left\lVert A \phi-b \right\rVert_2^2.
\end{equation}

\subsection{Time-stepping Method}
\vspace{0.3\baselineskip}
For a time-dependent PDE written in semi-discrete form:
\begin{equation}
\frac{\partial \phi}{\partial t}+\mathcal{N}(\phi)=0,
\end{equation}
the discrete state $\phi^n$ is advanced to $\phi^{n+1}$ using the unified scheme:
\begin{equation}
\frac{\phi^{n+1}-\phi^{n}}{\Delta t}
+\alpha\,\mathcal{N}(\phi^{n})
+(1-\alpha)\,\mathcal{N}(\phi^{n+1})
=0,\quad\alpha\in\{0,0.5,1\}.
\label{eq:theta_scheme}
\end{equation}
where $\Delta t$ is the time-step size and $\mathcal{N}(\cdot)$ denotes the spatial operator after finite-difference discretization (including boundary treatments). 
The parameter $\alpha$ controls the degree of implicitness: $\alpha=1$ recovers the forward (explicit) Euler scheme, $\alpha=0$ corresponds to the backward (implicit) Euler scheme, and $\alpha=0.5$ yields the Crank-Nicolson discretization~\citep{crank1947practical}.

Instead of advancing Eq.~\eqref{eq:theta_scheme} with a conventional time-integration implementation, such as explicit marching or implicit
updates that require solving linear/nonlinear systems at each step, FDTO
recasts each time-step update as a local optimization problem. At the
\(n\)-th time step, the state from the previous step is used to initialize
the optimization variable,
\[
\phi_0^{n+1} = \phi^{n}.
\]
The next state is then obtained by approximately minimizing the discrete
PDE residual over a single time step:
\begin{equation}
\phi^{n+1}
\approx
\arg\min_{\phi}\;
\mathcal{L}_{\mathrm{PDE}}^{n}(\phi^{n}),
\label{eq:time_marching_loss}
\end{equation}
where \(\mathcal{L}_{\mathrm{PDE}}^{n}\) denotes the residual loss associated
with the \(n\)-th time-step discretization.
Within each time step, this local optimization problem is solved
iteratively, with the number of inner iterations denoted by
\(K_{max}\). Starting from \(\phi_0^{n}\), the optimizer updates the
candidate state according to
\begin{equation}
\phi_{k+1}^{n}
=
\mathcal{O}\!\left(
\phi_k^{n},
\nabla_{\phi}\mathcal{L}_{\mathrm{PDE}}^{n}
\left(\phi_k^{n}\right),
\eta
\right),
 k=0,\ldots,K_{max}-1,
\label{eq:inner_optimizer_update}
\end{equation}
where \(\mathcal{O}\) denotes the optimizer at the \(n\)-th time step,
\(\eta\) is the learning rate, and the gradient is taken with respect to
the current candidate state \(\phi_k^{n}\). Boundary conditions are imposed
during the construction of the candidate state before evaluating the
residual loss.
Upon completion of the inner optimization loop, the optimized state is accepted as
\(\phi_{K_{max}}^{n}\)
and is then used to initialize the optimization at the next time step.
This procedure preserves the time-marching structure while replacing the
conventional algebraic solve at each step with a residual-minimization
process.

This local-in-time formulation improves the conditioning of the
optimization problem by restricting each solve to a single time-step
update, rather than optimizing over the entire time horizon simultaneously.
Moreover, the finite-difference discretization provides a naturally scaled
residual representation, which helps balance the gradients across different
state components, such as velocity and pressure, and thereby promotes stable
convergence when using first-order or quasi-Newton optimizers.

\subsection{Optimization Algorithm}
Algorithm~\ref{alg:fdto} summarizes the procedure for advancing the solution within the FDTO framework and serves as the basis for the algorithmic workflow discussed in this subsection.

The current solution $\phi^n$ serves as the initialization for the inner iterations, during which discrete residuals are evaluated and minimized. All spatial operators are applied locally through a finite-difference operator $\mathcal{N}$, while boundary conditions are enforced to residual evaluation to ensure valid stencil access near domain boundaries.

Grid-related quantities, including coordinate metrics and connectivity information, are precomputed and remain fixed throughout the optimization. As a result, each inner iteration operates only on the evolving flow variables, enabling efficient and stable loss and gradient evaluations. The optimization terminates when the maximum number of inner iterations is reached.

Once the inner loop optimization terminates, the updated state is accepted as $\phi^{n+1}$ and the algorithm proceeds to the next time step. The optimizer state is reset before each new step to avoid carrying over step-specific history. This local-in-time strategy circumvents global matrix assembly and inversion, while preserving the stencil structure and numerical stability of the underlying finite-difference discretization.
\begin{algorithm}[t]
\caption{Finite difference and time-stepping optimization}
\label{alg:fdto}
\DontPrintSemicolon
\KwIn{Initial state field $\big[\bm{u}^0,{p}^0\big]$, where 
\(\bm{u}^0=(u^0,v^0)\); learning rate $\eta$; max time steps $N_{t}$; max inner iterations $K_{\max}$}
\KwOut{Optimized field $\phi^{N_{t}}$; trajectory $\{\phi^n\}_{n=0}^{N_{t}}$}  
\BlankLine
\textbf{Initialize:} $\phi^0\leftarrow \big[\bm{u}^0,{p}^0\big]$; \;
\textbf{Define:} FDM operator $\mathcal{N}$ and Optimizer $\mathcal{O}$\;
\BlankLine
\For{\(n=0\) to \(N_t-1\)}{
$\phi^{n+1}_{0} \leftarrow \phi^{n}$ \;
\(\mathcal{O} \leftarrow \operatorname{CreateOptimizer}(\phi^{n+1}_{0},  \eta)\) \;
\For{$k=0$ \KwTo $K_{max}-1$}{
\(\tilde{\phi}^n|_{\partial\Omega} \leftarrow \operatorname{ImposeBC}(\phi^n)\)  \;
\(\tilde{\phi}^{n+1}_k|_{\partial\Omega} \leftarrow \operatorname{ImposeBC}(\phi^{n+1}_k)\)  \;
\({\phi}^\ast \leftarrow \mathcal{N}(\tilde{\phi}^{n+1}_k,\tilde{\phi}^n)\) \;
$\mathcal{L}_{\mathrm{PDE}}({\phi}^\ast) \leftarrow \frac{1}{N_h} \sum_{i=1}^{N_h} \left\lVert r_{\mathrm{PDE}, i}({\phi}^\ast) \right\rVert_2^2$\;
\({\phi}^{n+1}_{k+1}
\leftarrow
\operatorname{\mathcal{O}}
({\phi}^\ast,
\nabla_{\phi}\mathcal{L}_{\mathrm{PDE}}({\phi}^\ast),\eta)\) \;
}
\(\phi^{n+1} \leftarrow \phi^{n+1}_{K_{\max}}\)
}
\KwRet{\(\phi^{N_{t}}\)}\;
\end{algorithm}

\begin{figure*}[!ht]
\centering
\includegraphics[width=0.8\textwidth]{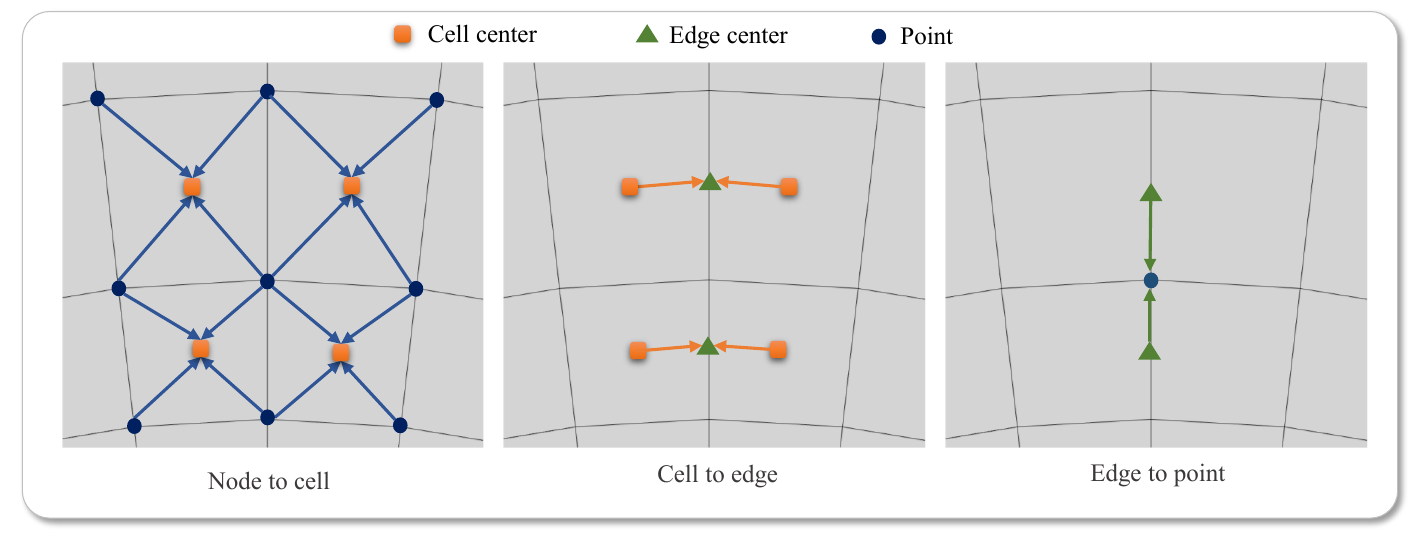}
\caption{Schematic of N-C-N averaging operator on body-fitted grids.}
\label{fig:ncn_smoothing}
\end{figure*}

Furthermore, an additional stabilization strategy is introduced to mitigate error accumulation and suppress pressure oscillations in wake-dominated airfoil flows. As illustrated in Fig.~\ref{fig:ncn_smoothing}, a node-to-cell-to-node (N-C-N) averaging operator is applied as a local smoothing on body-fitted structured grids. First, cell-centered values are computed as the arithmetic mean of the four corner nodal values for each adjacent cell. Next, these cell-center values are pairwise averaged along the computational $\xi$ direction to obtain two intermediate face-centered values. Finally, the two face-centered values are averaged along the $\eta$ direction to update the nodal value. This purely discrete, locality-preserving procedure damps high-frequency numerical noise that may accumulate during optimization-driven time marching, thereby improving the stability of pressure reconstruction in regions with strong shear and sharp gradients. In practice, the N-C-N smoothing is applied after each outer time-marching update to enhance interface consistency and suppress spurious oscillations on curvilinear body-fitted grids.

\section{Results and Discussions}
\label{sec:res}
This section assesses the proposed FDTO solver on a set of representative cases spanning different grid configurations, boundary conditions, and physical models, with primary emphasis on incompressible Navier--Stokes equations. The incompressible tests include lid-driven cavity benchmarks, external airfoil aerodynamics on body-fitted grids, and a cylinder case on a multi-block structured grid. Since ODIL baselines are not yet available for time-dependent PDEs in the current setting, the diffusion equation and a nonlinear flow-mixing problem are included as additional evaluations to demonstrate broader applicability. Reference solutions are taken from analytical expressions when available, and otherwise obtained from a conventional CFD solver (e.g., COMSOL). The relative $L_2$ error, defined as ${\|\phi_{\mathrm{opt}} - \phi_{\mathrm{ref}}\|_{L_2} \,/\, \|\phi_{\mathrm{ref}}\|_{L_2}}$
is used for all subsequent error calculations. Here, $\phi_{\mathrm{opt}}$ denotes the optimized solution obtained from the proposed method. The reference solution $\phi_{\mathrm{ref}}$ represents the ground-truth solution used for comparison. All experiments were conducted on a server equipped with six NVIDIA GeForce RTX 5090 GPUs, each providing 32 GB of VRAM, with Driver Version 570.144, CUDA Version 12.8, and PyTorch 2.8.0+cu128. 
\begin{table*}[!ht]
\centering
\setlength{\tabcolsep}{2pt}
\caption{Case configurations for evaluating the FDTO solver.}
\begin{tabular}{lccccclcc}
\hline
\multicolumn{1}{c}{Equation Type} & Geometry          & Grid type   & Grid size & Optimizer & Epoch & Delta t & Inner iterations & Learning rate \\ \hline
\multicolumn{1}{c}{Diffusion}      & Square            & Uniform     & 101$\times$101   & L-BFGS    & 100   & 1/100   & 100             & 1.0           \\ 
\multicolumn{1}{c}{Flow Mixing}    & Square            & Uniform     & 256$\times$256   & L-BFGS    & 256   & 1/64    & 100             & 1.0           \\ 
Incompressible                    & Cavity & Uniform     & 201$\times$201   & L-BFGS    & 1000  & 1/10    & 100             & 1.0           \\ 
Navier--Stokes                    & Airfoil           &Curvilinear & 399$\times$150   & SOAP      & 1000  & 1/100    & 2000            & 5e-4          \\  
                                  & Cylinder          & Multi-block & 29801     & SOAP      & 3000  & 1/10    & 2000            & 5e-4          \\ \hline
\end{tabular}
\label{tab:case_config}
\end{table*}

Tab.~\ref{tab:case_config} summarizes the case configurations used to validate the proposed solver. The test set includes incompressible Navier--Stokes cases covering both internal and external flows (lid-driven cavity, airfoil, and cylinder), together with diffusion and flow-mixing problem on a square domain. Three grid types are considered: a uniform Cartesian grid for the square and cavity cases, a curvilinear body-fitted grid for the airfoil configuration, and a multi-block grid for the cylinder case. L-BFGS is used for the uniform-grid problems, whereas the airfoil and cylinder cases adopt the SOAP optimizer with a smaller learning rate. The time-step size $\Delta t$, epoch budget, and inner iterations are selected case by case to reflect differences in conditioning and geometric complexity, enabling a systematic assessment of robustness and accuracy across physical models, geometries, and grid configurations.

\subsection{Lid-driven Cavity}
\begin{figure}[!ht]
\centering
\includegraphics[width=0.4\textwidth]{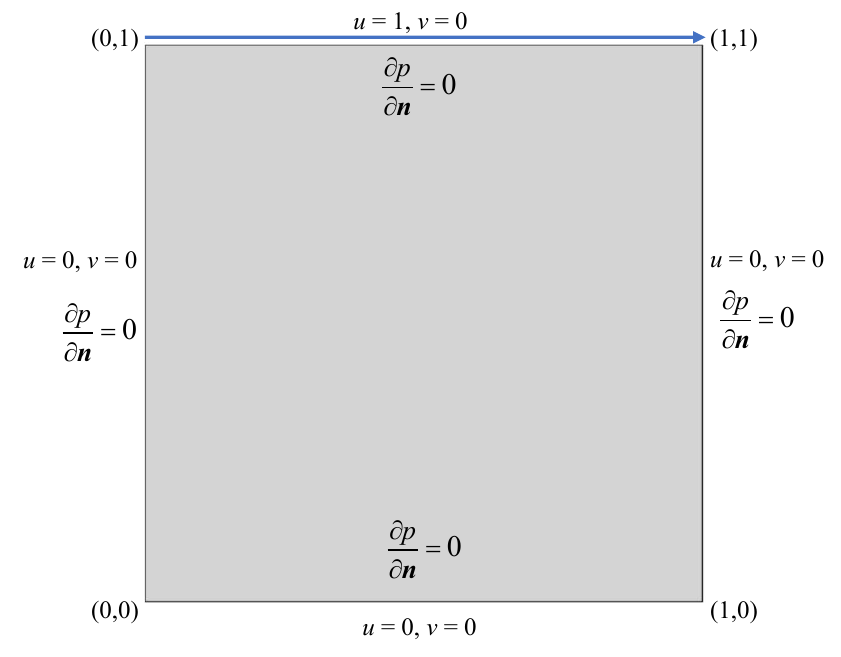}
\caption{Boundary Conditions for the lid-driven cavity (Uniform).}
\label{fig:BC_cavity}
\end{figure}

To begin the incompressible Navier--Stokes evaluation with a well-established internal-flow benchmark, the lid-driven cavity flow is considered. This canonical test enables a clear assessment of numerical accuracy, and high-fidelity reference data are available~\citep{ghia1982high}.
The physical domain is the unit square $[0,1]\times[0,1]$ and is discretized on a structured $201\times201$ grid. 
As shown in Fig.~\ref{fig:BC_cavity}, no-slip velocity conditions are imposed on the stationary walls, while the top lid moves with a prescribed tangential velocity $(u,v)=(1,0)$. 
For pressure, a homogeneous Neumann condition $\partial p/\partial n=0$ is applied on all boundaries. The performance of FDTO is compared with ODIL-SGR~\citep{cao2025overcoming} and representative PINN-based numerical differential (PINN-ND) solvers, such as FDGN and generative finite-volume graph network (Gen-FVGN)~\citep{li2025learning}. FDGN encodes finite-difference stencils into graph neural network message-passing operations to approximate the discrete differential operators, while Gen-FVGN leverages a generative graph neural network to predict finite-volume fluxes on unstructured meshes, enabling mesh-agnostic flow reconstruction.

\begin{figure*}[!ht]
\centering
\includegraphics[width=1.0\textwidth]{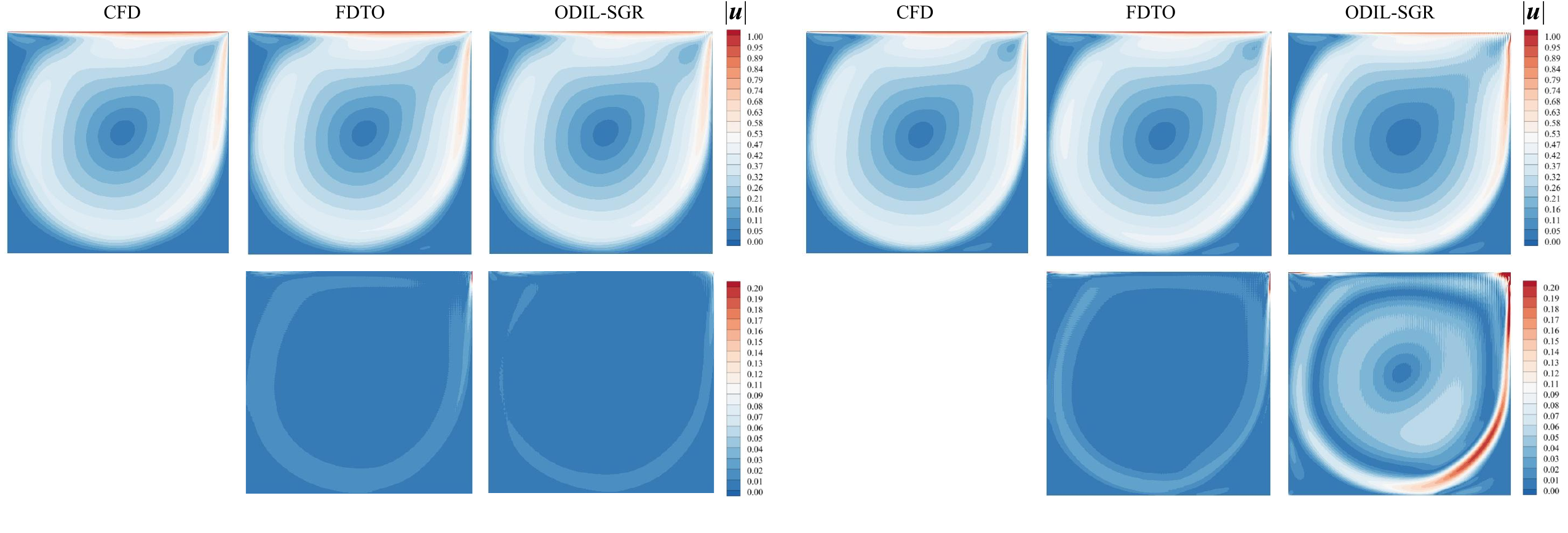}
\caption{Comparison of CFD, FDTO, and ODIL-SGR for the lid-driven cavity at $\mathrm{Re}=2500$ (left) and $\mathrm{Re}=5000$ (right)}, showing velocity fields and absolute error maps on a $201\times201$ grid.
\label{fig:cavity_Re2500_and_Re5000}
\end{figure*}

\begin{figure*}[!ht]
    \centering
    \includegraphics[width=0.7\textwidth]{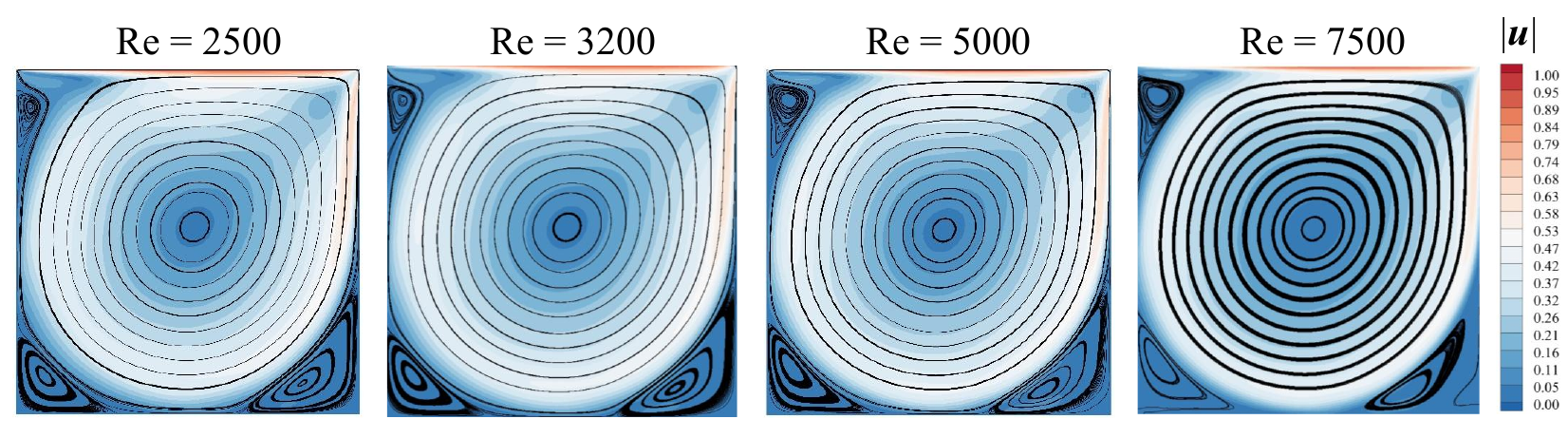}
    \caption{FDTO-optimized streamlines for lid-driven cavity flow from $\mathrm{Re}=2500$ to $\mathrm{Re}=7500$} on a $201\times201$ grid.
    \label{fig:cavity_streamline}
    \end{figure*}

Fig.~\ref{fig:cavity_Re2500_and_Re5000} illustrates the predicted velocity fields and the corresponding error maps at $\mathrm{Re}=2500$ and $\mathrm{Re}=5000$. ODIL-SGR achieves errors comparable to (and slightly lower than) FDTO at $\mathrm{Re}=2500$, but its accuracy degrades markedly at $\mathrm{Re}=5000$, where the errors become substantially larger than FDTO. By contrast, FDTO maintains consistently low error levels at both Reynolds numbers, indicating improved robustness as convection effects intensify with increasing Reynolds numbers.
The streamline visualizations in Fig.~\ref{fig:cavity_streamline} further corroborate the field-level comparison. FDTO reliably reproduces the expected vortex topology from $\mathrm{Re}=2500$ to $\mathrm{Re}=7500$, including the emergence of the characteristic multi-vortex structure at higher Reynolds numbers. 

\begin{figure*}[!ht]
\centering
\includegraphics[width=0.6\textwidth]{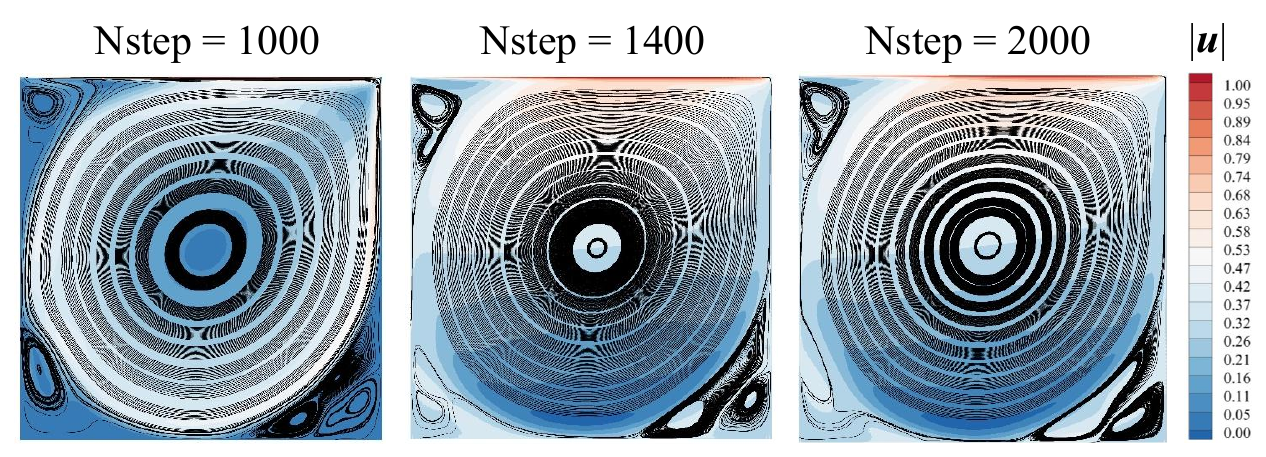}
\caption{FDTO-optimized lid-driven cavity streamlines at $\mathrm{Re}=10000$ on a $201\times201$ grid for different time steps ($N_{\mathrm{step}}=1000,1400,2000$).}
\label{fig:cavity_Re10000}
\end{figure*}

To probe a more challenging regime, the $\mathrm{Re}=10000$ case is initialized from the converged solution at $\mathrm{Re}=3200$. Time marching is then performed with $\Delta t=0.05$. Fig.~\ref{fig:cavity_Re10000} shows instantaneous streamlines at representative time instances, illustrating the temporal evolution of the vortex structures.

\begin{figure*}[!ht]
\centering
\includegraphics[width=0.8\textwidth]{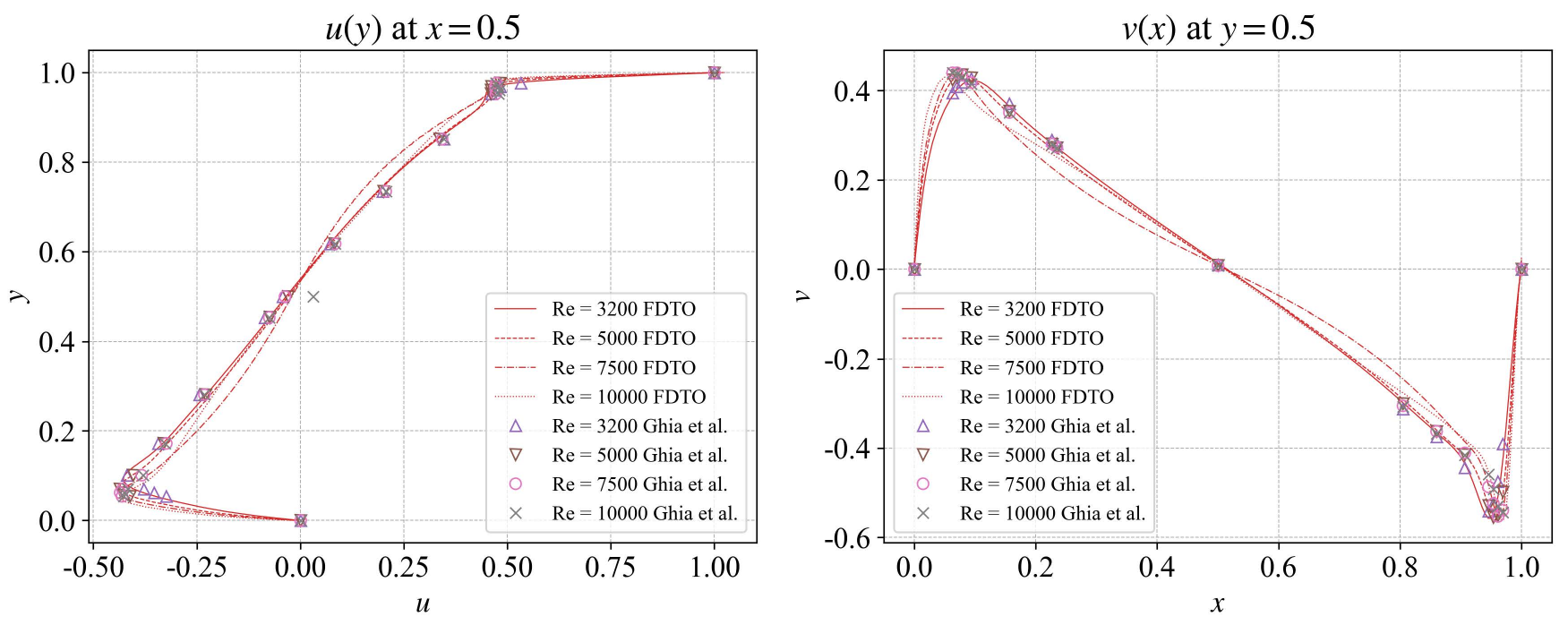}
\caption{Comparisons of horizontal velocity along the vertical centerline (left) and  vertical velocity along the horizontal centerline (right) for the time-averaged lid-driven cavity flows at $\mathrm{Re}=2500$} on a $201\times201$ grid.
\label{fig:cavity_center-verlocity}
\end{figure*}

\begin{figure*}[!ht]
\centering
\includegraphics[width=0.8\textwidth]{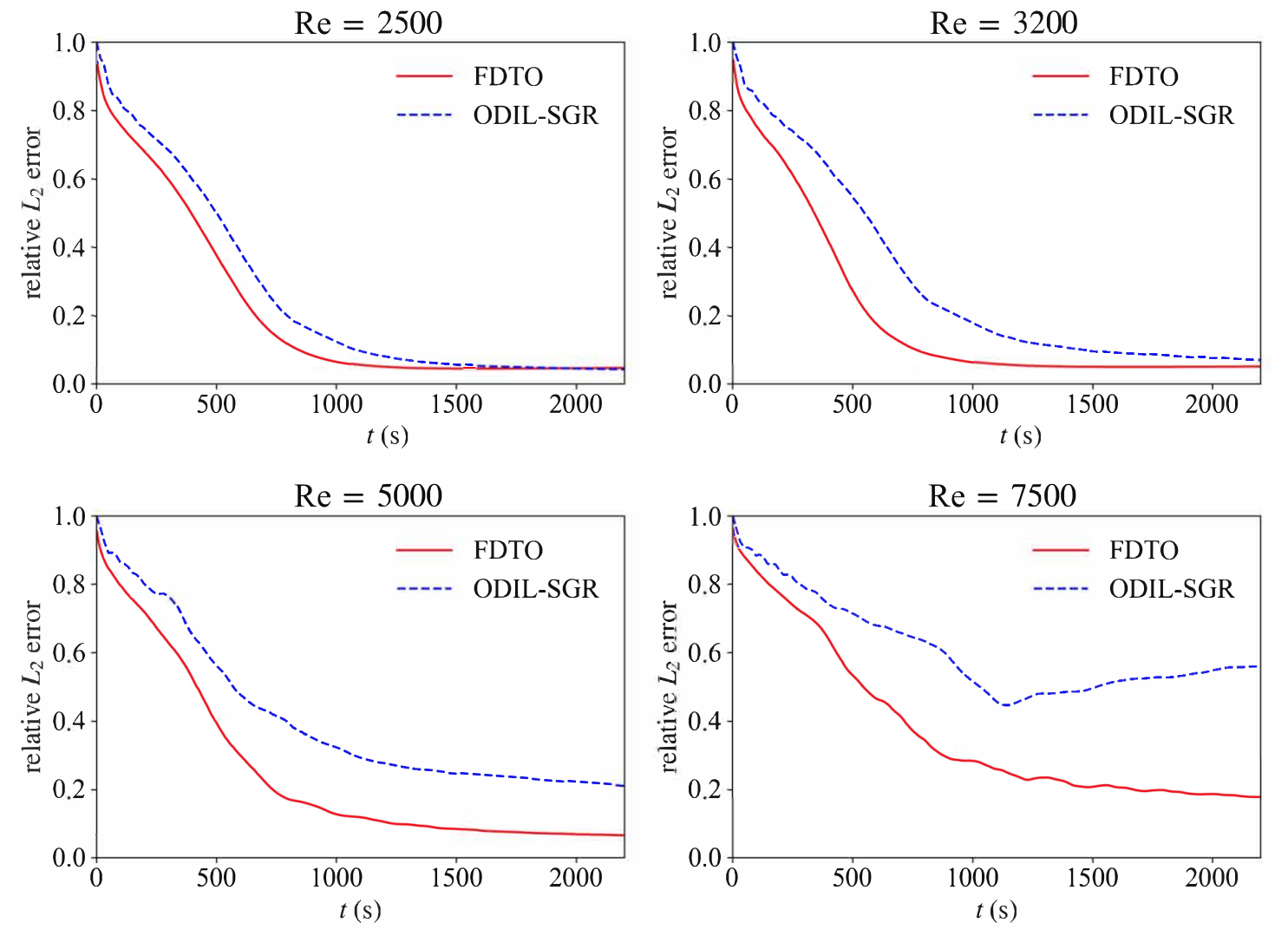}
\caption{Convergence comparison of FDTO and ODIL-SGR for lid-driven cavity flow on a $201\times201$ grid: relative $L_2$ error versus runtime at $\mathrm{Re}=2500$, 3200, 5000, and 7500.}
\label{fig:cavity_vs_SGR}
\end{figure*}

Beyond qualitative visualization, Fig.~\ref{fig:cavity_center-verlocity} reports representative centerline profiles, including velocity and pressure distributions along the cavity centerlines, together with the benchmark data of Ghia et al.~\citep{ghia1982high}. These profiles provide a concise quantitative check of the flow structure and confirm that FDTO reproduces the expected near-wall behavior and overall recirculation patterns in the time-dependent case.
Fig.~\ref{fig:cavity_vs_SGR} summarizes the convergence histories. Across Reynolds numbers, FDTO exhibits faster error reduction and consistently lower relative $L_2$ errors than ODIL-SGR, with the performance gap widening as $\mathrm{Re}$ increases and the optimization landscape becomes more nonlinear. Consistent with these trends, Tab.~\ref{tab:cavity-acc} reports the minimum achieved error and the corresponding runtime, highlighting the efficiency advantage of FDTO at higher Reynolds numbers. Notably, at $\mathrm{Re}=7500$, ODIL-SGR fails to converge to a stable solution, whereas FDTO still reaches a low-error state within a finite runtime.
\begin{table*}[!ht]
\centering
\setlength{\tabcolsep}{2pt}
\caption{Comparison of minimum error and convergence time for FDTO and ODIL-SGR in lid-driven cavity flow($201\times201$).}
\begin{tabular}{ccccccccc}
\hline
\multirow{2}{*}{Method}  & \multicolumn{2}{c}{$\mathrm{Re}=2500$} & \multicolumn{2}{c}{$\mathrm{Re}=3200$} & \multicolumn{2}{c}{$\mathrm{Re}=5000$} & \multicolumn{2}{c}{$\mathrm{Re}=7500$} \\ \cline{2-9} 
& Relative $L_2$ error   & $t(\mathrm{s})$  & Relative $L_2$ error   & $t(\mathrm{s})$  & Relative $L_2$ error   & $t(\mathrm{s})$  & Relative $L_2$ error   & $t(\mathrm{s})$  \\ \hline
FDTO                         & 0.05             & 1193    & 0.05             & 1409    & 0.06             & 2291    & 0.12             & 4510    \\ 
ODIL-SGR                     & 0.05             & 1689    & 0.06             & 2452    & 0.20             & 2445    & 0.57             & -    \\ \hline
\end{tabular}
\label{tab:cavity-acc}
\end{table*}

\begin{figure}[!ht]
\centering
\includegraphics[width=0.48\textwidth]{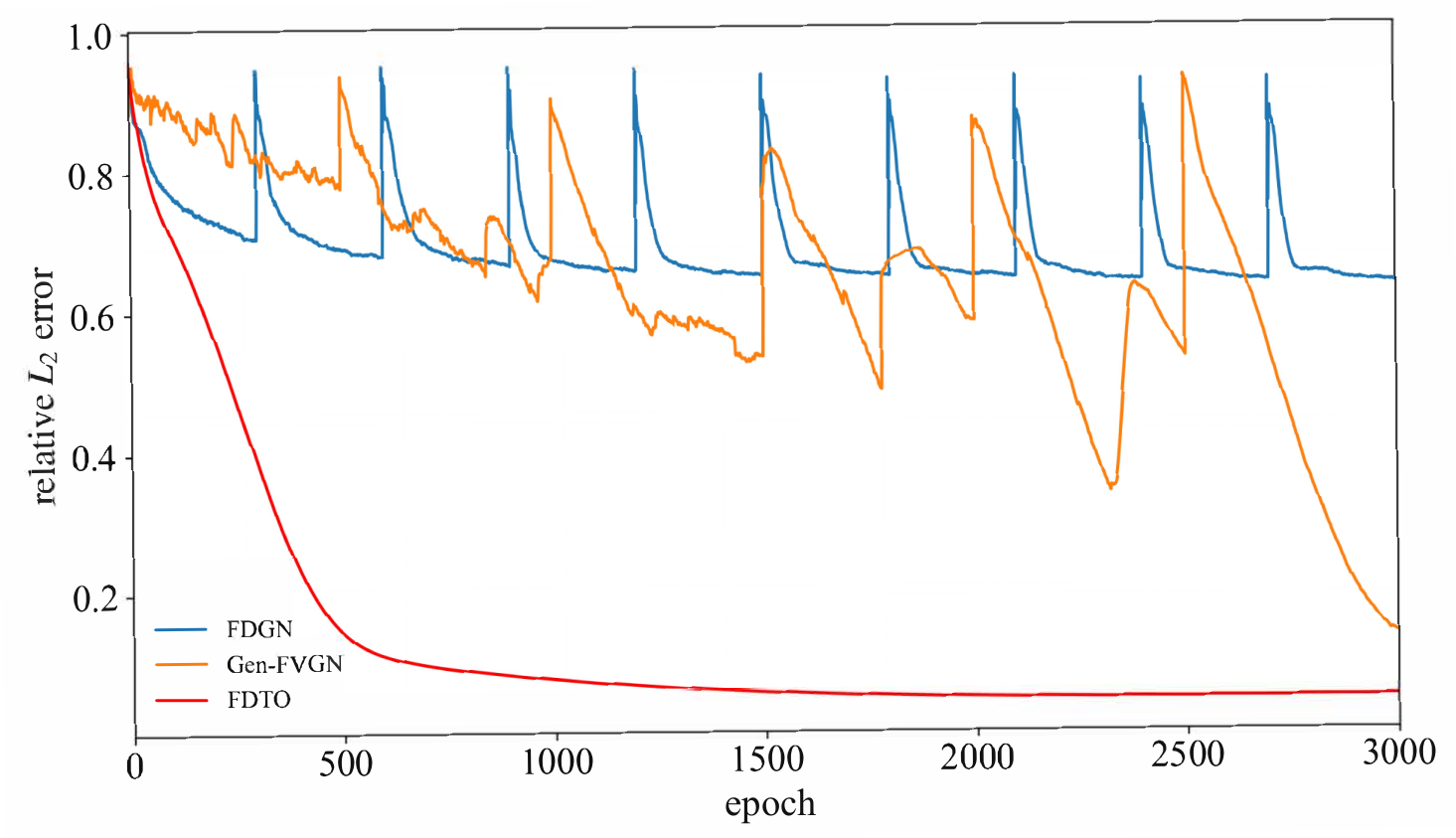}
\caption{Convergence comparison on lid-driven cavity flow at $\mathrm{Re}=2500$} on a $201\times201$ grid: relative $L_2$ error versus epoch for FDTO, FDGN, and Gen-FVGN.
\label{fig:cavity_FDX_fig}
\end{figure}

Finally, FDTO is compared with representative PINN-ND solvers by assessing practical efficiency (per-epoch runtime and GPU memory usage) together with convergence stability. The convergence curves in Fig.~\ref{fig:cavity_FDX_fig} highlight the stability of the proposed framework. FDTO exhibits smooth, nearly monotonic error decay, whereas FDGN and Gen-FVGN converge more slowly and show more pronounced oscillations, suggesting reduced optimization stability. From an efficiency standpoint, Tab.~\ref{tab:cavity_FDX_tab} further shows that FDTO achieves a substantially smaller GPU memory footprint than Gen-FVGN while maintaining competitive runtime, indicating a favorable trade-off between memory usage and computational speed. This trend is consistent with the core design of FDTO, where the unknown fields are optimized directly on discrete degrees of freedom under physics constraints rather than being represented by a global neural parameterization. By avoiding backpropagation through deep network surrogates and their associated intermediate activations, FDTO reduces memory overhead and maintains stable coupled updates of velocity and pressure under challenging dynamics.
\begin{table}[!ht]
\centering
\setlength{\tabcolsep}{1pt}
\caption{Comparison of GPU memory usage and time costs per epoch for different methods in lid-driven cavity flow($201\times201$, $\mathrm{Re}=2500$)}.
\begin{tabular}{ccc}
\hline
Method & GPU memory (MB)         & Time cost per epoch (s)         \\ \hline
FDGN                    & 4072                  & 2.52                            \\ 
Gen-FVGN                & 7611                  & 2.22                            \\ 
FDTO                    & 710                   & 1.02                            \\ \hline
\end{tabular}
\label{tab:cavity_FDX_tab}
\end{table}
\subsection{AirFoil Flow}
\begin{figure}[!ht]
\centering
\includegraphics[width=0.48\textwidth]{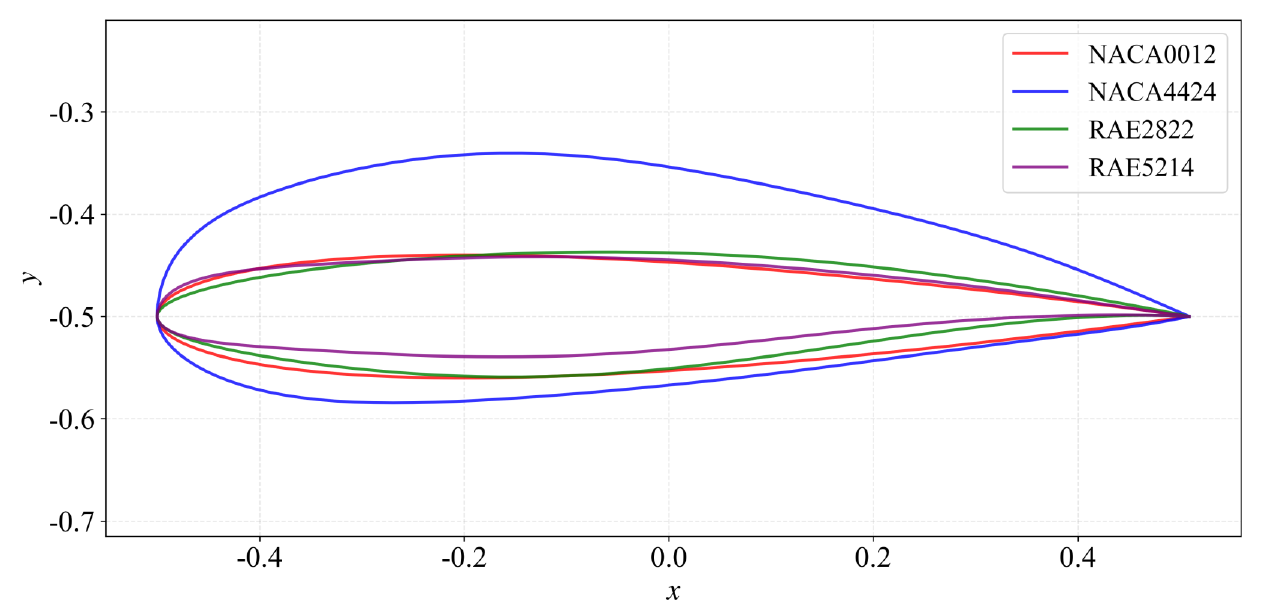}
\caption{Representative airfoil geometries (NACA0012, NACA4424, RAE2822, and RAE5214).}
\label{fig:geo_airfoil}
\end{figure}

\begin{figure*}[!ht]
\centering
\includegraphics[width=0.8\textwidth]{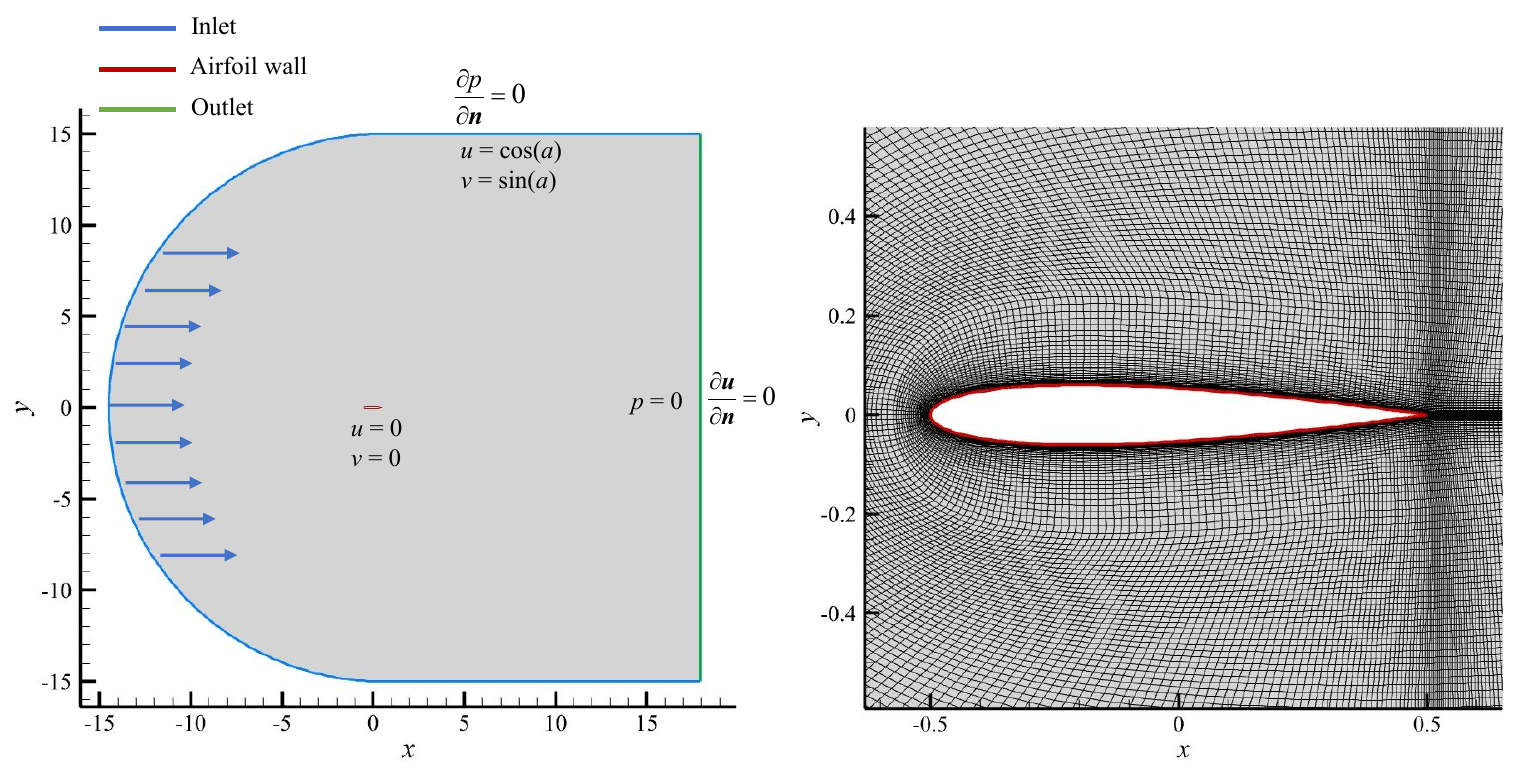}
\caption{Computational domain, boundary conditions (left), and body-fitted structured grid (right) for airfoil flows (curvilinear).}
\label{fig:BC_airfoil}
\end{figure*}
When extending the evaluation from canonical internal flows to external aerodynamic configurations, pressure reconstruction becomes more challenging on body-fitted grids, especially in wake-dominated regions where strong shear and sharp gradients may trigger oscillations during optimization-based time stepping. This subsection focuses on engineering-relevant airfoil cases. The goal is to assess geometric adaptability on body-fitted structured grids, robustness of pressure reconstruction in convection-dominated regimes, and preservation of key aerodynamic quantities of interest. Four representative airfoils (NACA0012, NACA4424, RAE2822, and RAE5214) are considered, as shown in Fig.~\ref{fig:geo_airfoil} and summarized in Tab.~\ref{tab:AirFoil}. Unless otherwise specified, all cases are computed on a $399\times150$ body-fitted structured grid (Fig.~\ref{fig:BC_airfoil}). The computational domain consists of an upstream semicircular far-field boundary and the top/bottom far-field boundaries, together with a downstream outlet boundary, the airfoil surface is treated as a solid wall.
On the far-field boundary, a uniform free stream with angle of attack $\alpha$ is prescribed for velocity, while a homogeneous Neumann condition is imposed for pressure:$(u,v)=(\cos\alpha,\ \sin\alpha), {\partial p}/{\partial n}=0$, where $n$ denotes the outward unit normal. On the outlet boundary, a reference pressure is prescribed and zero-normal-gradient conditions are applied to the velocity: $p=0$, ${\partial u}$/${\partial n}$=0 and ${\partial v}$/${\partial n}$=0.
On the airfoil wall, a no-slip condition is imposed: $u=0$ and $v=0$.
All boundary specifications are enforced through the ghost-node linear extrapolation closure described in Sec.~\ref{sec:FDTO_framework}, ensuring stencil-consistent gradient evaluation near the body-fitted boundary and stable pressure reconstruction in the wake region.
\begin{table}[!ht]
\centering
\setlength{\tabcolsep}{20pt}
\caption{Airfoil test-case configurations (Reynolds number and angle of attack).}
\begin{tabular}{ccc}
\hline
Cases & Re   & $\alpha$   \\ \hline
NACA0012 & 100  & $15^\circ$ \\ 
NACA4424 & 200  & $11^\circ$ \\ 
RAE2822  & 500  & $7^\circ$  \\ 
RAE5214  & 1000 & $3^\circ$  \\ \hline
\end{tabular}
\label{tab:AirFoil}
\end{table}

\begin{figure*}[!ht]
\centering
\includegraphics[width=0.98\textwidth]{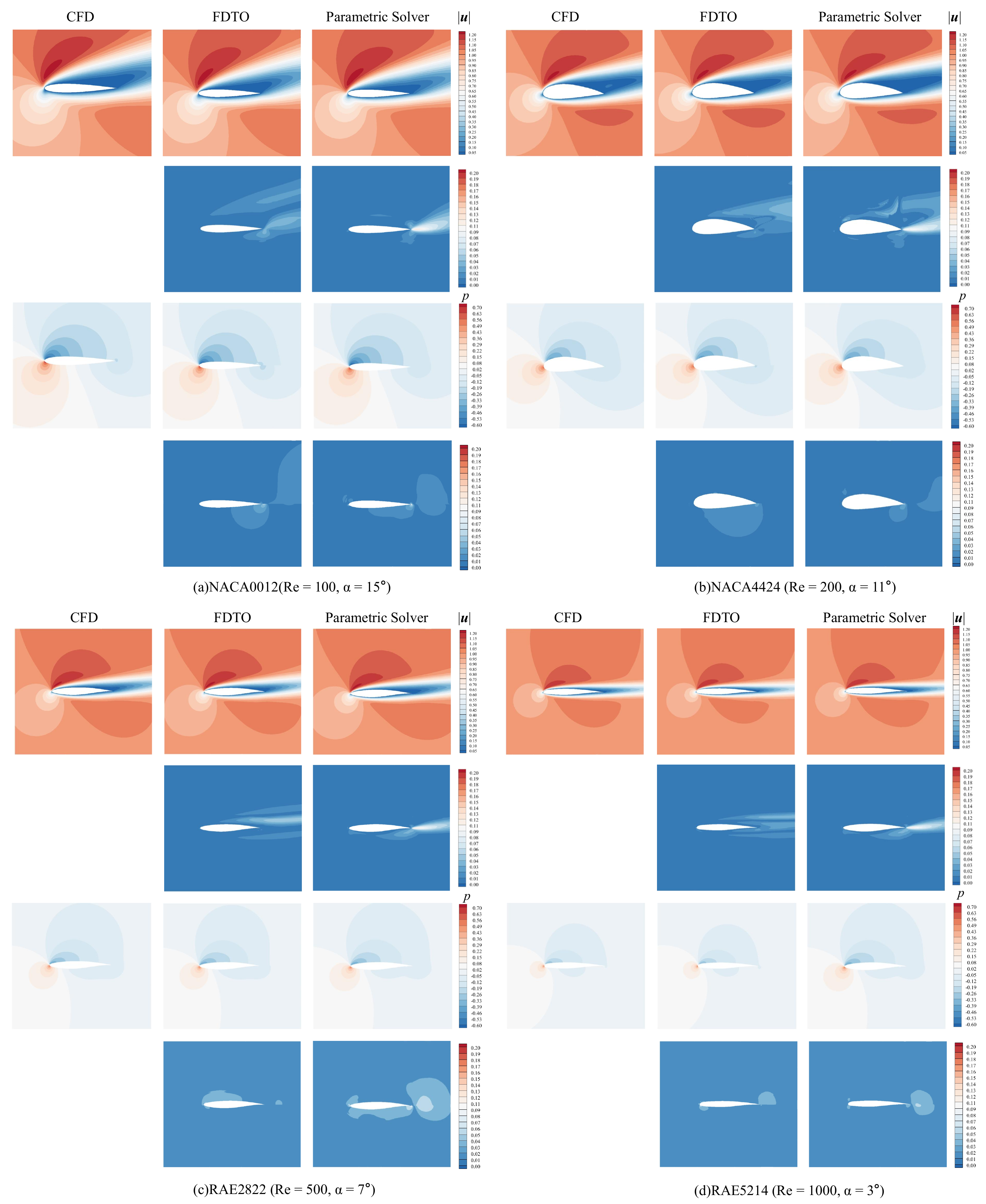}
\caption{Comparison of airfoil flow fields and absolute error maps.For each case, the columns (left to right) show the reference CFD solution, FDTO prediction, and the parametric solver prediction.
The rows (top to bottom) report the velocity-magnitude contours, the absolute velocity error, the pressure contours, and the absolute pressure error.}
\label{fig:AirFoil}
\end{figure*}

Fig.~\ref{fig:AirFoil} (a-d) illustrates the velocity and pressure fields for four representative airfoil cases (NACA0012, NACA4424, RAE2822, and RAE5214), comparing the reference CFD solutions with the predictions of FDTO and the parametric solver~\citep{cao2025surrogate}. For each case, velocity-magnitude contours and absolute velocity errors are reported, together with pressure contours and absolute pressure errors. Across all configurations, FDTO reproduces the global flow topology around the airfoil and yields small errors that remain localized to sharp-gradient regions (e.g., near the trailing edge and the near wake). This behavior supports geometric adaptability and flow-field fidelity on body-fitted grids. By comparison, the parametric solver exhibits larger and more spatially distributed errors, particularly in the downstream wake, indicating reduced robustness in shear-dominated regions.

\begin{figure*}[!ht]
\centering
\includegraphics[width=0.8\textwidth]{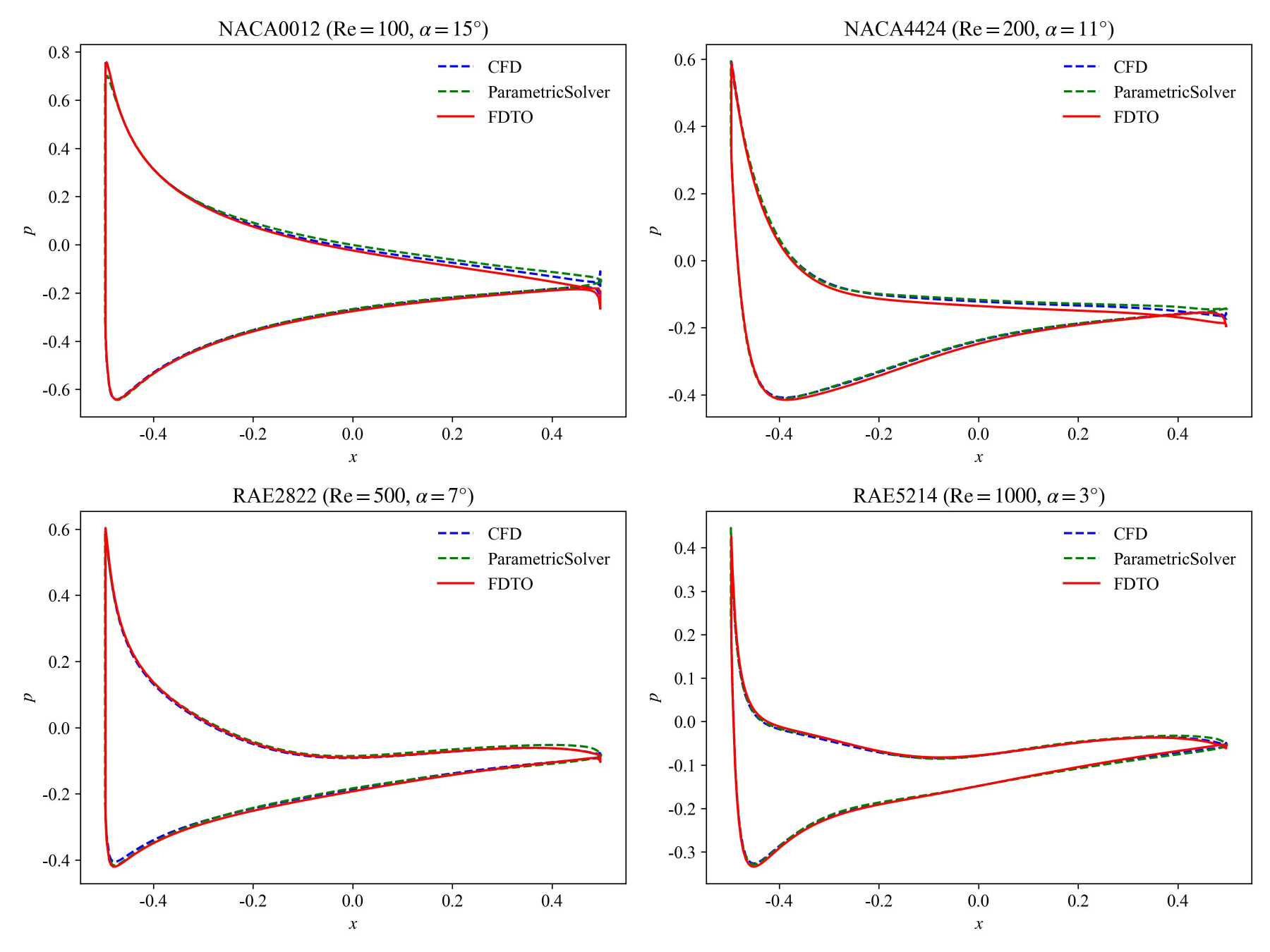}
\caption{Comparison of surface pressure distributions: CFD, FDTO, and the parametric solver.
Top-left: NACA0012($\mathrm{Re}=100$, $\alpha=15^\circ$); top-right: NACA4424($\mathrm{Re}=200$, $\alpha=11^\circ$); bottom-left: RAE2822($\mathrm{Re}=500$, $\alpha=7^\circ$); bottom-right: RAE5214($\mathrm{Re}=1000$}, $\alpha=3^\circ$).
\label{fig:surface pressure}
\end{figure*}

Beyond flow-field contours, wall-pressure consistency is assessed using the surface pressure distribution, a sensitive diagnostic for external aerodynamics. Fig.~\ref{fig:surface pressure} shows that FDTO closely matches the CFD reference across all airfoils, capturing both the leading-edge suction peak and the downstream pressure recovery for symmetric and cambered geometries. By contrast, the parametric solver exhibits more pronounced deviations near strong-gradient regions, consistent with the error patterns observed in the wake and trailing-edge vicinity.

Following the qualitative field comparisons in Fig.~\ref{fig:AirFoil}, quantitative assessments are further reported in terms of solution errors and integrated aerodynamic loads.
Tab.~\ref{tab:airfoil_error} reports the relative $L_2$ errors of $|\bm{u}|$ and $p$ across all cases. FDTO attains consistently lower velocity errors while maintaining competitive pressure accuracy, with more pronounced gains in configurations featuring stronger nonlinear wake interactions (e.g., RAE2822). 
Beyond field-level metrics, Tab.~\ref{tab:airfoil_coeff} compares the lift and drag coefficients among CFD, the parametric solver, and FDTO. Across different Reynolds numbers and angles of attack, FDTO remains in close agreement with the CFD references and exhibits reduced deviations-particularly in drag for the more nonlinear cases-indicating that improved flow and pressure reconstruction translates into reliable integrated aerodynamic loads.
\begin{table*}[!ht]
\centering
\setlength{\tabcolsep}{10pt}
\caption{Relative $L_2$ errors of $|\bm{u}|$ and $p$ for airfoil cases: FDTO versus the parametric solver.}
\begin{tabular}{cccc}
\hline
Cases & Method & $|\bm{u}|$ Relative $L_2$ Error & $p$ Relative $L_2$ Error \\ \hline
\multirow{2}{*}{\begin{tabular}[c]{@{}c@{}}NACA0012 ($\mathrm{Re}=100$, $\alpha=15^\circ$)\end{tabular}} & ParametricSolver & 0.016 & 0.060 \\ 
 & FDTO & 0.011 & 0.068 \\ 
\multirow{2}{*}{\begin{tabular}[c]{@{}c@{}}NACA4424 ($\mathrm{Re}=200$, $\alpha=11^\circ$)\end{tabular}} & ParametricSolver & 0.016 & 0.049 \\ 
 & FDTO & 0.012 & 0.069 \\ 
\multirow{2}{*}{\begin{tabular}[c]{@{}c@{}}RAE2822 ($\mathrm{Re}=500$, $\alpha=7^\circ$)\end{tabular}} & ParametricSolver & 0.014 & 0.054 \\ 
 & FDTO & 0.009 & 0.033 \\ 
\multirow{2}{*}{\begin{tabular}[c]{@{}c@{}}RAE5214 ($\mathrm{Re}=1000$, $\alpha=3^\circ$)\end{tabular}} & ParametricSolver & 0.009 & 0.042 \\ 
 & FDTO & 0.008 & 0.049 \\ \hline
\end{tabular}
\label{tab:airfoil_error}
\end{table*}
\begin{table*}[!ht]
\centering
\setlength{\tabcolsep}{23pt}
\caption{Comparison of lift and drag coefficients ($C_l$, $C_d$) for airfoil cases: CFD, FDTO, and the parametric solver.}
\begin{tabular}{cccc}
\hline
Cases  & Method           & $C_l$     & $C_d$     \\ \hline
\multirow{3}{*}{\begin{tabular}[c]{@{}c@{}} \end{tabular}} & CFD              & 0.6626 & 0.5036 \\ 
 NACA0012($\mathrm{Re}=100$, $\alpha=15^\circ$) & ParametricSolver & 0.6727 & 0.4826 \\ 
 & FDTO             & 0.6498 & 0.5118 \\ 
\multirow{3}{*}{\begin{tabular}[c]{@{}c@{}}\end{tabular}} & CFD              & 0.3223 & 0.3821 \\ 
 NACA4424 ($\mathrm{Re}=200$, $\alpha=11^\circ$) & ParametricSolver & 0.3498 & 0.3770 \\ 
 & FDTO             & 0.3391 & 0.3947 \\ 
\multirow{3}{*}{\begin{tabular}[c]{@{}c@{}}\end{tabular}}   & CFD              & 0.3581 & 0.1849 \\ 
 RAE2822 ($\mathrm{Re}=500$, $\alpha=7^\circ$) & ParametricSolver & 0.3741 & 0.1833 \\ 
 & FDTO             & 0.3742 & 0.1889 \\ 
\multirow{3}{*}{\begin{tabular}[c]{@{}c@{}}\end{tabular}}  & CFD              & 0.2191 & 0.1176 \\ 
 RAE5214 ($\mathrm{Re}=1000$, $\alpha=3^\circ$) & ParametricSolver & 0.2222 & 0.1187 \\ 
 & FDTO             & 0.2244 & 0.1203 \\ \hline
\end{tabular}
\label{tab:airfoil_coeff}
\end{table*}
\subsection{Cylinder}
\begin{figure*}[!ht]
\centering
\includegraphics[width=1.0\textwidth]{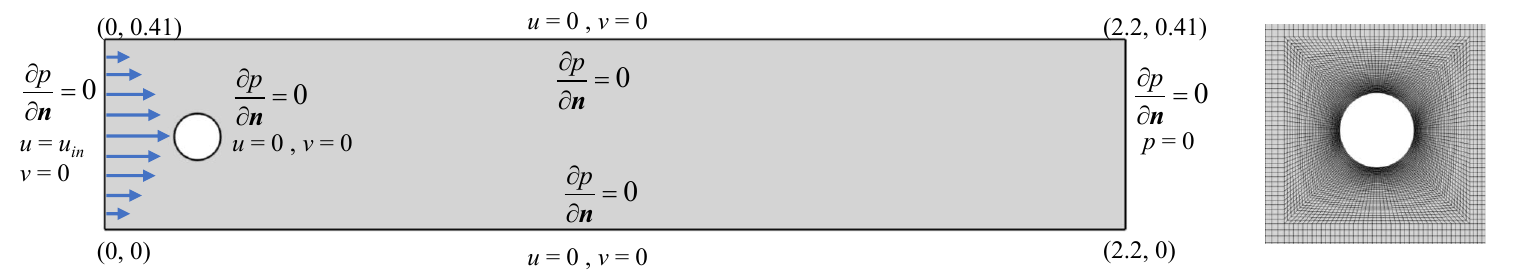}
\caption{Cylinder-in-channel configuration: boundary conditions (left) and the body-fitted multi-block structured grid (right).}
\label{fig:Cylinder-grid}
\end{figure*}
To assess geometric adaptability on multi-block structured grids, incompressible flow around a cylinder in a channel is considered. In this setting, block interfaces and inter-block connectivity may introduce numerical artifacts unless the discrete state remains coherent across blocks. As shown in Fig.~\ref{fig:Cylinder-grid}, the computational domain is a rectangle $(x,y)\in[0,2.2]\times[0,0.41]$ with a circular cylinder embedded near the inlet. The grid is partitioned into multiple structured blocks. A body-fitted, locally refined block surrounds the cylinder to resolve near-wall gradients and separated shear layers, while an elongated downstream block uses stretched spacing to efficiently capture wake development. The resulting multi-block grid contains approximately $2.98\times10^4$ points, together with connectors associated with block interfaces.
The left boundary is prescribed as a uniform inflow: $u=u_{\mathrm{in}}$, $v=0$, ${\partial p}$/${\partial n}$=0.
On the cylinder surface and the channel walls (top and bottom), no-slip conditions are enforced:$u=0$, $v=0$, ${\partial p}$/${\partial n}$=0, where $n$ denotes the outward unit normal. At the right boundary, a reference pressure is prescribed as $p=0$, and a compatible outflow pressure closure is applied via ghost-node extrapolation consistent with the near-boundary stencils. All boundary conditions and inter-block interface continuity are enforced using the unified ghost-node linear extrapolation and interface-consistent transfer operators described in Sec.~\ref{sec:FDTO_framework}, helping maintain cross-block coherence and suppress spurious oscillations near block connectors.

\begin{figure*}[!ht]
\centering
\includegraphics[width=1.0\textwidth]{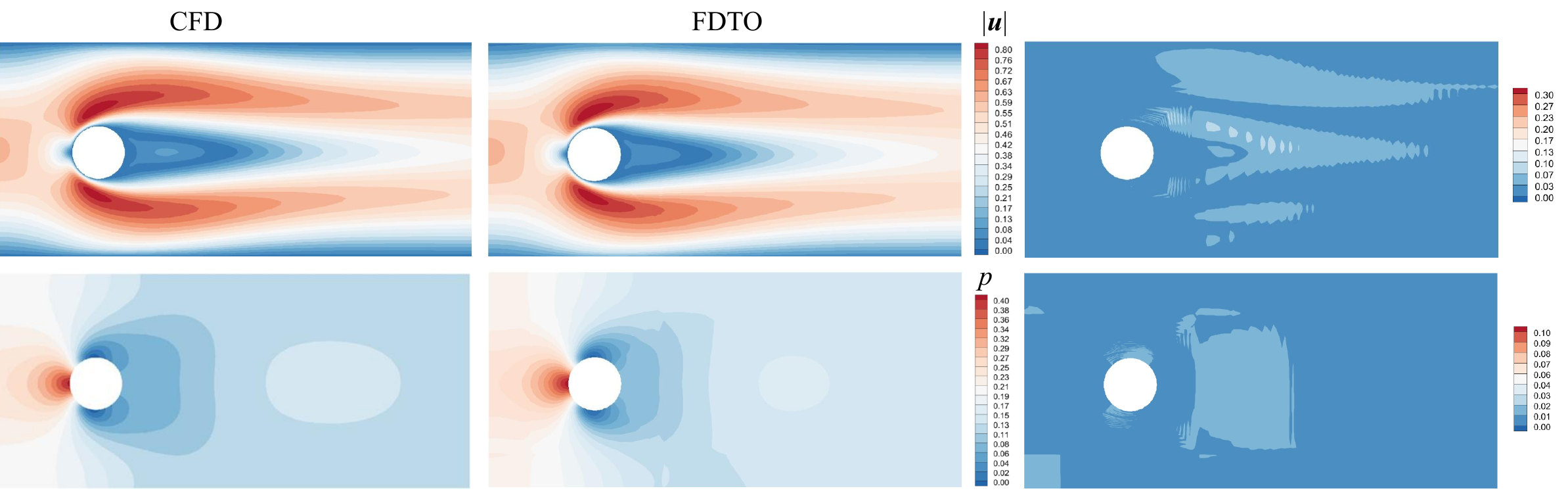}
\caption{Comparison of FDTO against CFD for cylinder flow.
Top row: velocity-magnitude and error; bottom row: pressure fields and error.}
\label{fig:Cylinder}
\end{figure*}

With the multi-block grid and boundary conditions specified above, solution quality and cross-block coherence are assessed on this external-flow case. Fig.~\ref{fig:Cylinder} compares the CFD reference solution with FDTO for the velocity magnitude $|\bm{u}|$ and pressure $p$, along with the corresponding absolute error maps. FDTO reproduces the dominant wake physics, including acceleration/deceleration around the cylinder, the low-speed recirculation region, and the downstream evolution of separated shear layers. Quantitatively, FDTO attains relative $L_2$ errors of $4.1651\times10^{-2}$ for $|\bm{u}|$ and $6.7093\times10^{-2}$ for $p$, demonstrating agreement with the CFD reference in both velocity and pressure fields. 
Importantly, the errors remain localized to gradient-dominated regions (e.g., near-wall pressure variations and the shear-layer/wake core) rather than forming bands aligned with block interfaces. This indicates that the proposed framework preserves cross-block field coherence and mitigates interface-related artifacts on multi-block grids.
\subsection{Diffusion Equation}
Accordingly, a nonlinear diffusion equation is considered as a representative parabolic PDE widely used to describe heat and scalar transport processes~\citep{cho2023separable}. The governing equation is given by
\begin{equation}
\partial_t u = \alpha \left(\|\nabla u\|^2 + u \Delta u \right),
\quad x \in \Omega, \; t \in \Gamma,
\end{equation}
subject to the initial and boundary conditions
\begin{equation}
u(x,0) = u_{\mathrm{ic}}(x), \quad x \in \Omega,
\end{equation}
\begin{equation}
u(x,t) = 0, \quad t \in \Gamma.
\end{equation}

The spatial domain is defined as $\Omega = [-1,1]^2$, and the temporal domain is $\Gamma = [0,1]$. The diffusion coefficient is set to $\alpha = 0.05$. The initial condition is constructed as a superposition of three Gaussian functions:
\begin{equation}
\begin{aligned}
u_{\mathrm{ic}}(x,y) = \;& 0.25 \exp \left[-10\{(x-0.2)^2 + (y-0.3)^2\}\right] \\
&+ 0.4 \exp \left[-15\{(x+0.1)^2 + (y+0.5)^2\}\right] \\
&+ 0.3 \exp \left[-20\{(x+0.5)^2 + y^2\}\right].
\end{aligned}
\end{equation}

\begin{figure*}[!ht]
\centering
\includegraphics[width=1.0\textwidth]{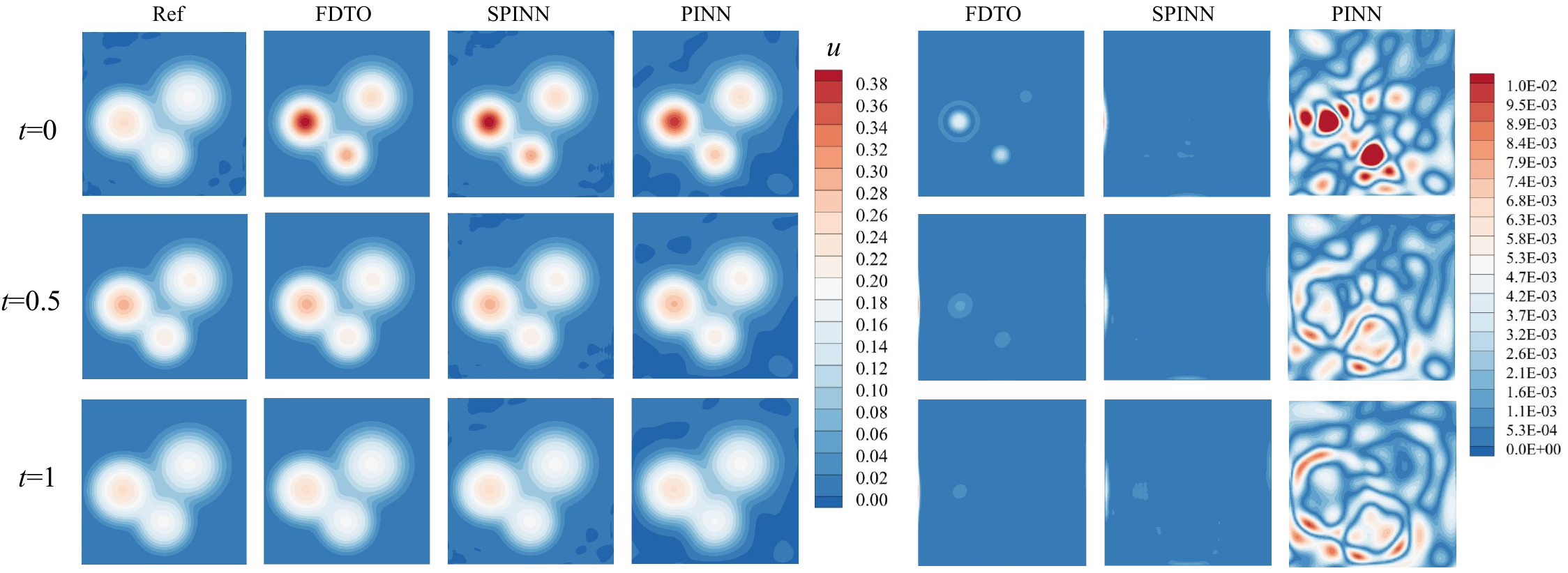}
\caption{Comparison of flow field (left) and absolute error (right) for  diffusion equation on a $101\times101$ grid.}
\label{fig:Diffusion}
\end{figure*}

A comparative study is conducted between the proposed FDTO solver, separable PINN (SPINN), and a standard PINN, using the reference solution reported in~\citep{cho2023separable} as the ground-truth baseline. Fig.~\ref{fig:Diffusion} presents the solution fields together with the corresponding absolute error maps on a $101\times101$ grid at $t=0$, $0.5$, and $1$. Both FDTO and SPINN closely reproduce the reference solution across all snapshots, with small errors that remain confined to localized regions. In contrast, the standard PINN exhibits noticeable deviations in the background region and near steep gradients, together with spatially diffuse error patterns. Consistent with these distributions, the PINN yields substantially larger errors at all time instances, indicating reduced fidelity in capturing the diffusion dynamics under the same evaluation setting. These results suggest that optimizing directly over discrete degrees of freedom allows FDTO to preserve diffusion dynamics with localized, physically consistent errors, while avoiding the diffuse error spread characteristic of a globally parameterized PINN.

Consistent with the trends observed in the solution and error contours, Tab.~\ref{tab:diffusion-error} provides a quantitative comparison of relative $L_2$ error and GPU memory consumption for the diffusion equation. FDTO maintains low errors at $t=0$, $0.5$, and $1$, while SPINN achieves comparable (slightly smaller) errors. From a computational perspective, SPINN is the most memory-efficient, requiring 590~MB of GPU memory. FDTO uses slightly more memory (636~MB) but remains lightweight and practical. In contrast, the PINN requires 8762 MB-over an order of magnitude higher than both FDTO and SPINN (approximately 13.8 times compared with FDTO). Taken together, these results indicate that FDTO achieves near-SPINN-level accuracy while maintaining low GPU memory usage, and substantially outperforms the PINN baseline in both accuracy and memory footprint on this diffusion test.
\begin{table}[!ht]
\centering
\setlength{\tabcolsep}{2pt}
\caption{Comparison of relative $L_2$ error and GPU memory for the diffusion equation on a $101\times101$ grid.}
\label{tab:diffusion-error}
\begin{tabular}{c ccc c}
\hline
\multirow{2}{*}{Method} 
& \multicolumn{3}{c}{Relative $L_2$ error}
& \multirow{2}{*}{GPU memory (MB)} \\ \cline{2-4}
& $t=0\,\mathrm{s}$   & $t=0.5\,\mathrm{s}$ & $t=1\,\mathrm{s}$ &  \\ \hline  
SPINN  & 0.005   & 0.004   & 0.003    & 590    \\ 
PINN   & 0.046   & 0.034   & 0.038    & 8762   \\ 
FDTO   & 0.004   & 0.005   & 0.005    & 636      \\ \hline  
\end{tabular}
\end{table}
\subsection{Flow-mixing Problem}
To further assess the proposed solver, a more challenging $(2+1)$-dimensional flow-mixing case is considered, representing the mixing of two fluids. Following the setup in SPINN~\citep{cho2023separable,chiu2022can}, the governing equation is given by
\begin{equation}
\begin{aligned}
\frac{\partial u(t,x,y)}{\partial t}
+ a\,\frac{\partial u(t,x,y)}{\partial x}
+ b\,\frac{\partial u(t,x,y)}{\partial y} &= 0,\\
t \in [0,4],\quad x \in [-4,4],\quad y \in [-4,4], &
\end{aligned}
\end{equation}
where
\begin{equation}
a(x,y) = -\frac{v_t}{v_{t,\max}}\frac{y}{r},
\end{equation}
\begin{equation}
b(x,y) = \frac{v_t}{v_{t,\max}}\frac{x}{r},
\end{equation}
\begin{equation}
v_t = \mathrm{sech}^2(r)\tanh(r),
\end{equation}
\begin{equation}
r = \sqrt{x^2 + y^2},
\end{equation}
with $v_{t,\max} = 0.385$. The spatial domain is $\Omega = [-4,4]\times[-4,4]$, and the temporal domain is $t \in [0,4]$.
This problem has an analytical solution:
\begin{equation}
u(t,x,y)
= -\tanh \left(
\frac{y}{2}\cos(\omega t) - \frac{x}{2}\sin(\omega t)
\right),
\end{equation}
where $\omega = v_t/({r\,v_{t,\max}})$. The initial and boundary conditions are extracted directly from the analytical solution.

Methods are evaluated on a structured $256\times256$ grid with the same temporal discretization as in the previous section. The governing equations and boundary conditions are specified consistently for FDTO, SPINN, and PINN. For a fair comparison, solution fields are sampled on an identical spatiotemporal grid, and the analytical solution is used as the reference.

\begin{figure*}[!ht]
\centering
\includegraphics[width=1.0\textwidth]{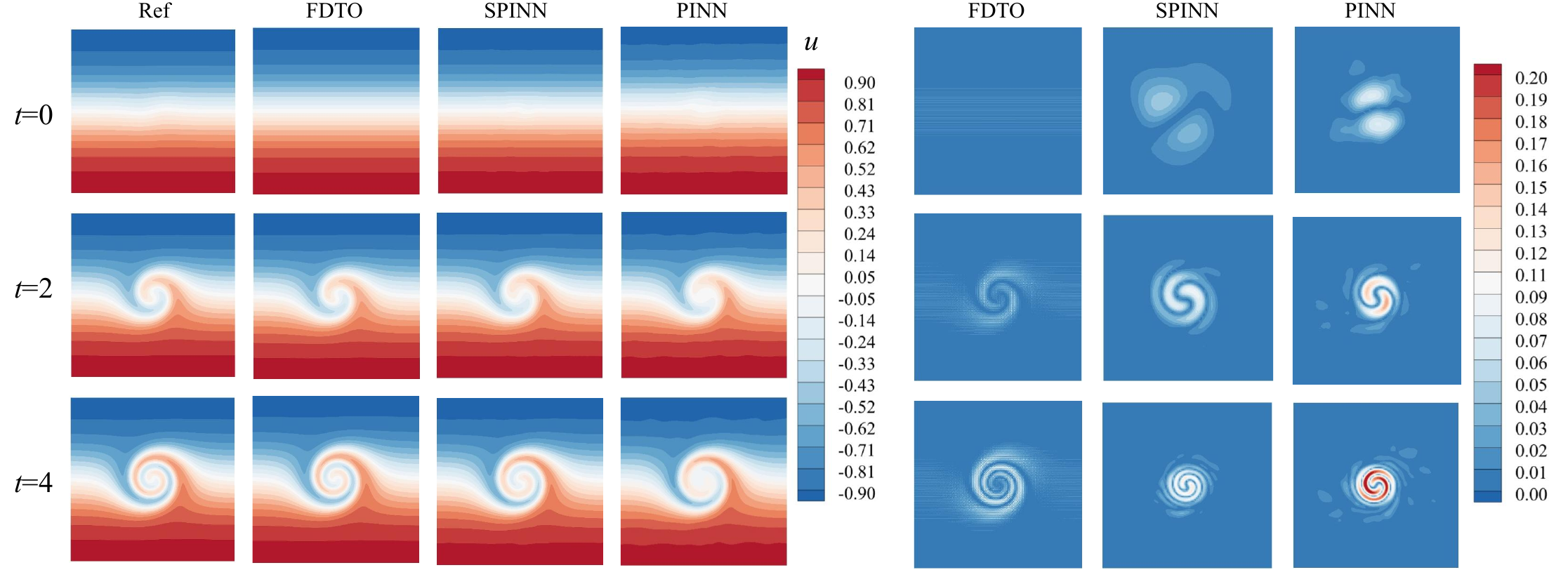}
\caption{Comparison of flow field (left) and absolute error (right) for flow-mixing problem on a $256\times256$ grid.}
\label{fig:FlowMix}
\end{figure*}

Fig.~\ref{fig:FlowMix} compares the solution fields and absolute error maps at $t=0$, $2$, and $4$ for a linear advection problem featuring strong rotational deformation and interface stretching. FDTO accurately captures the spiral mixing structures and remains in close agreement with the reference across all snapshots, with errors primarily confined to the thin spiral interface where gradients are strongest and negligible in the bulk region. SPINN reproduces the main mixing pattern but exhibits slight smoothing in strongly deformed regions, leading to moderate interface-localized errors. By comparison, PINN shows pronounced numerical diffusion and interface distortion, especially at later times, with substantially larger and more scattered errors around the vortex core and along the mixing interface.

To complement the qualitative comparisons in the solution and error contours, Tab.~\ref{tab:FlowMix} summarizes the quantitative results, reporting the relative $L_2$ errors together with GPU memory consumption. FDTO achieves the lowest error at $t=0$ and maintains its advantage as the mixing evolves. At $t=2$, FDTO attains a relative $L_2$ error of 0.014, which is nearly three times smaller than SPINN and about five times smaller than PINN. Similar trends persist at $t=4$. In terms of resource usage, SPINN is the most memory efficient, while FDTO requires slightly more GPU memory but remains lightweight and practical. In contrast, PINN consumes roughly twice the GPU memory of FDTO. The proposed solver achieves a favorable balance between solution fidelity and memory efficiency.

\begin{table}[!ht]
\centering
\setlength{\tabcolsep}{2pt}
\caption{Comparison of relative $L_2$ error and GPU memory for the flow-mixing problem on a $256\times256$ grid.}
\begin{tabular}{c ccc c}
\hline
\multirow{2}{*}{Method} 
& \multicolumn{3}{c}{Relative $L_2$ error} 
& \multirow{2}{*}{GPU memory (MB)} \\ \cline{2-4}
& $t=0\,\mathrm{s}$ & $t=2\,\mathrm{s}$ & $t=4\,\mathrm{s}$ &  \\ \hline
SPINN & 0.010 & 0.040 & 0.045 & 570  \\ 
PINN  & 0.022 & 0.068 & 0.066 & 1590 \\ 
FDTO  & 0.009 & 0.014 & 0.023 & 670  \\ \hline
\end{tabular}
\label{tab:FlowMix}
\end{table}
\subsection{Ablation Studies}
\subsubsection{Convergence Comparison}
\begin{figure}[!ht]
\centering
\includegraphics[width=0.48\textwidth]{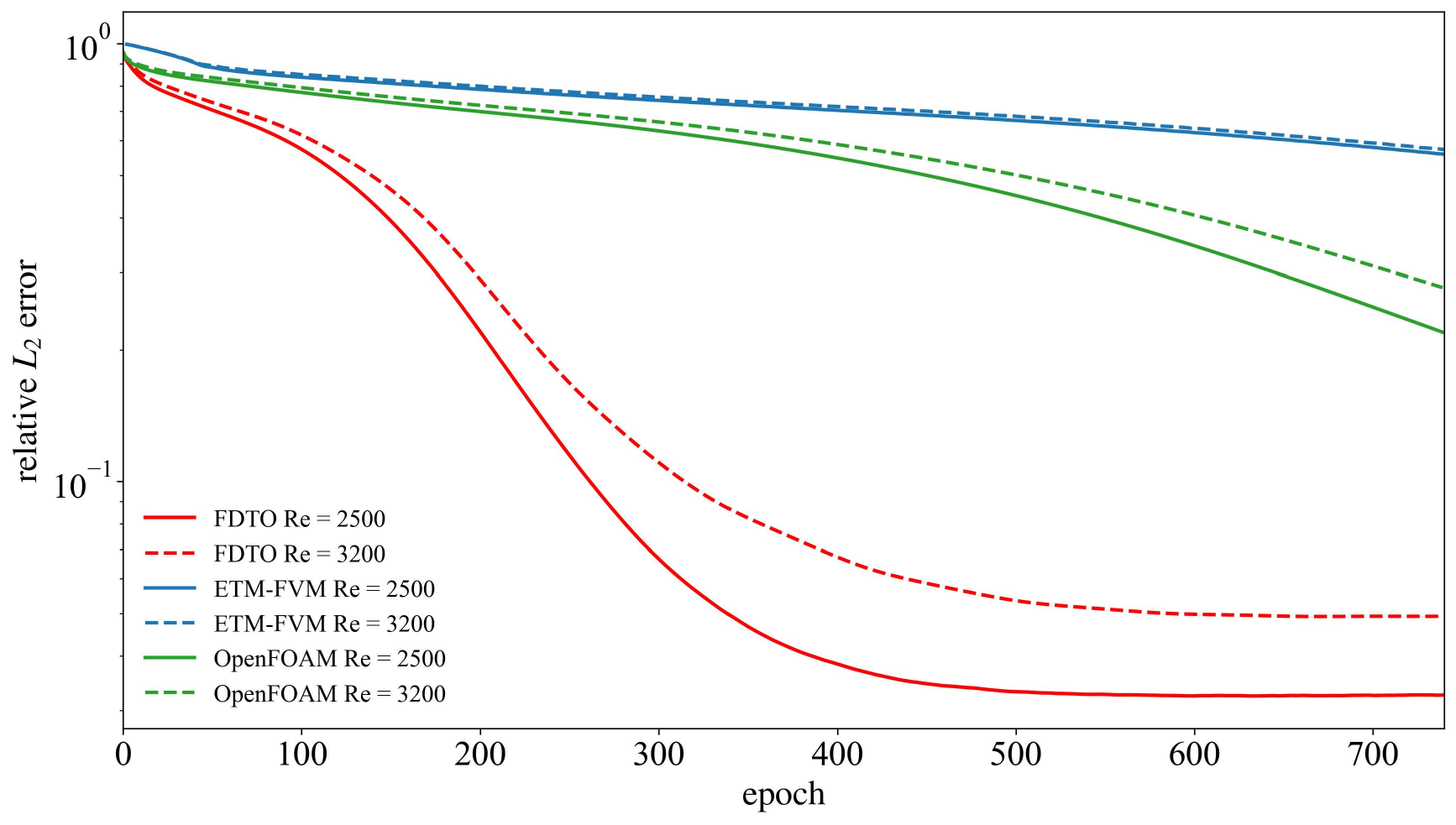}
\caption{Comparison of relative $L_2$ error versus epoch for FDTO, ETM-FVM (explicit), and OpenFOAM pimpleFoam (implicit, backward Euler scheme) on a $201\times201$ lid-driven cavity grid at $\mathrm{Re}=2500$ and $\mathrm{Re}=3200$}.
\label{fig:FDTO_vs_CFD}
\end{figure}

Beyond optimization-based baselines, FDTO is further compared with both a conventional explicit time-marching finite-volume (ETM-FVM) solver and an implicit time-marching finite-volume solver~\citep{fripp_fv_cfd_matlab}. The implicit baseline employs OpenFOAM's \texttt{pimpleFoam} solver with a backward time-stepping scheme, which is representative of commonly used implicit CFD solvers. Fig.~\ref{fig:FDTO_vs_CFD} presents the convergence histories of the proposed FDTO solver, the ETM-FVM solver, and the implicit OpenFOAM solver for the lid-driven cavity at $\mathrm{Re}=2500$ and $\mathrm{Re}=3200$. In both cases, the ETM-FVM advances the flow field using an explicit update with a CFL-limited time step chosen to be the largest empirically stable value, while the OpenFOAM \texttt{pimpleFoam} solver advances the solution using an implicit backward scheme with a comparatively larger time step; FDTO, in contrast, performs optimization-driven time marching by directly minimizing the discrete residual.

The results demonstrate that FDTO reaches a low-error plateau within roughly 300--400 epochs with smooth and nearly monotonic decay, achieving a relative $L_2$ error on the order of $10^{-2}$ at $\mathrm{Re}=2500$ and slightly above $10^{-2}$ at $\mathrm{Re}=3200$. The implicit OpenFOAM solver exhibits moderate convergence, reducing the error to approximately $2 \times 10^{-1}$ at $\mathrm{Re}=2500$ and $3 \times 10^{-1}$ at $\mathrm{Re}=3200$ after 750 epochs, which is notably better than the explicit ETM-FVM but still falls substantially short of the accuracy achieved by FDTO. The ETM-FVM shows the slowest convergence among the three methods, with the relative $L_2$ error remaining above $5 \times 10^{-1}$ across the entire training budget. As the Reynolds number increases, the convergence turning point of FDTO is mildly delayed and the plateau error increases slightly for all three methods. Nevertheless, FDTO converges markedly faster than both the explicit ETM-FVM and the implicit OpenFOAM solver across all cases. In particular, the advantage of FDTO over the implicit solver---which already benefits from larger, CFL-free time steps---indicates that the efficiency gain of FDTO goes beyond merely relaxing the CFL constraint, and stems from the optimization-driven nature of the time-marching scheme that directly targets residual minimization rather than relying on fixed-coefficient linear solves at each step.
\subsubsection{Inner Iterations Study}
\begin{figure*}[!t]
\centering
\includegraphics[width=0.96\textwidth]{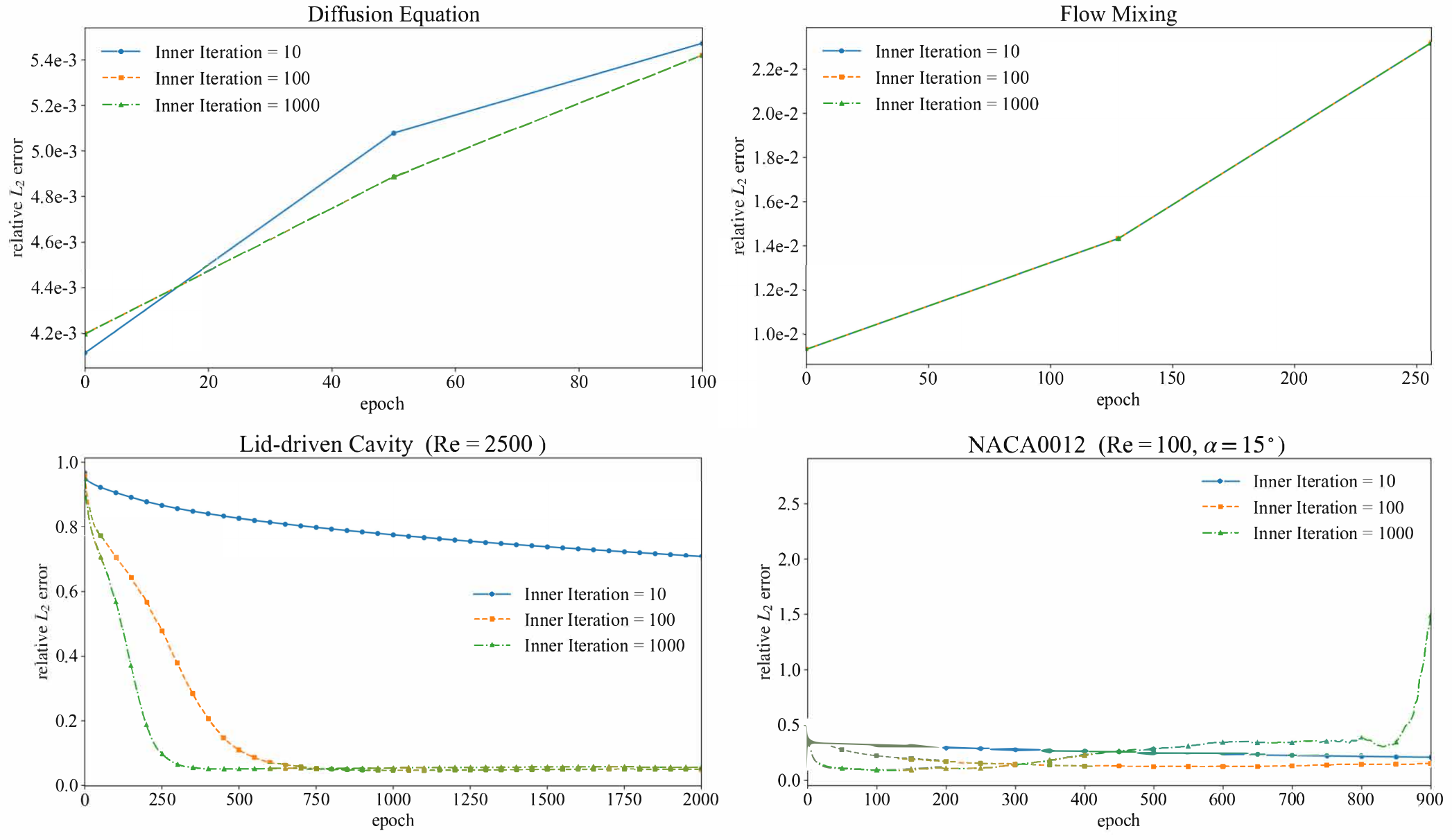}
\caption{Comparison of inner iterations on convergence across four cases: diffusion equation, flow-mixing problem, lid-driven cavity ($\mathrm{Re}=2500$}) and NACA0012 ($\mathrm{Re}=100$, $\alpha=15^\circ$).
\label{fig:iteration_compare}
\end{figure*}

This subsubsection investigates the sensitivity of FDTO to the number of inner iterations per epoch(i.e., per outer time-marching update). As shown in Fig.~\ref{fig:iteration_compare}, the impact of the inner iterations budget is strongly case-dependent. For the diffusion and flow-mixing problem, the three curves nearly overlap: increasing inner iterations from 10 to 1000 yields only marginal changes, and the relative $L_2$ error follows a similar trend as epochs progress (from nearly $4\times10^{-3}$ to $5\times10^{-3}$ for diffusion, and from $10^{-2}$ to $2.3\times10^{-2}$ for flow mixing), indicating that the overall error is primarily governed by the outer update and discretization rather than insufficient inner solves. 

In contrast, the lid-driven cavity case is highly sensitive to inner iterations: with only 10 inner iterations, the error decreases slowly and remains large even after 2000 epochs, whereas 100 or 1000 inner iterations produce a rapid error drop and reach a low-error plateau around $5\times10^{-2}$, with 1000 iterations converging noticeably faster in the early stage but attaining a similar final level as 100 iterations (diminishing returns). 

For the NACA0012 case, 100 inner iterations improve convergence over 10 iterations and reaches a lower steady error level. In contrast, 1000 inner iterations becomes unstable: after an initially rapid decrease, the error gradually increases and eventually diverges. This behavior suggests that an overly aggressive inner iterations budget can destabilize the coupled outer update in a more nonlinear, geometry-influenced setting. 

These results indicate a practical trade-off. Increasing the inner iterations budget can substantially accelerate convergence when inner under-optimization is the main bottleneck (e.g., the lid-driven cavity case), but it offers limited benefit when convergence is dominated by the outer update (diffusion/mixing) and may even reduce robustness (NACA0012). A moderate choice (e.g., 100) provides a favorable balance among accuracy, computational cost, and stability across the tested cases.
\subsubsection{Effect of Time-step Size}
\begin{figure*}[!ht]
\centering
\includegraphics[width=0.96\textwidth]{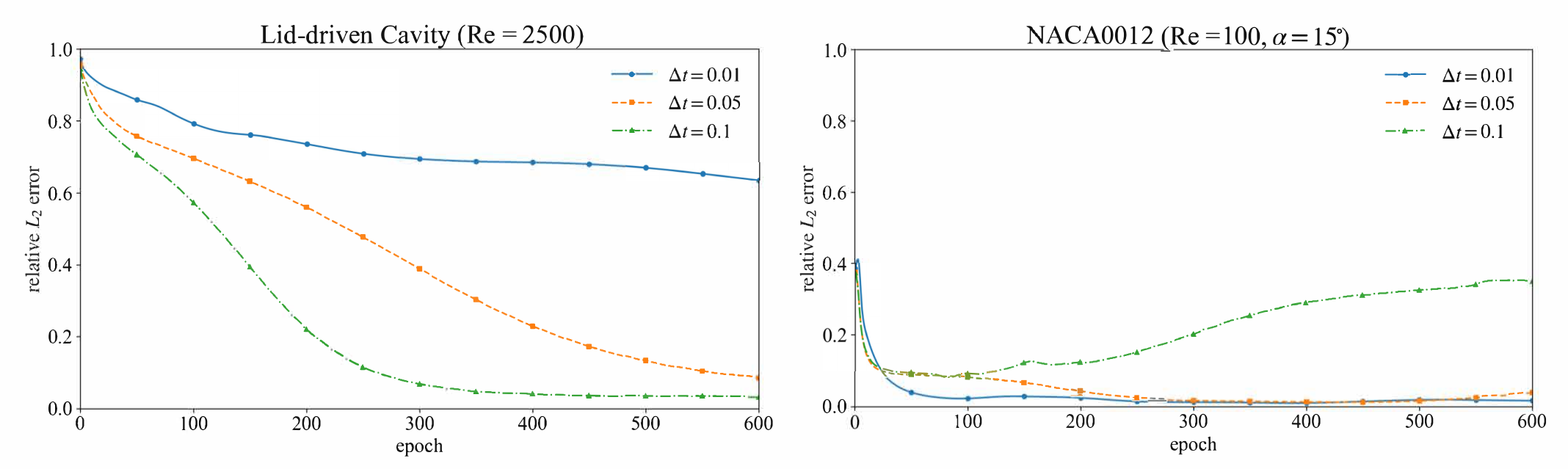}
\caption{Effect of time-stepping choice ($\Delta t$=0.1, 0.05, 0.01) on lid-driven cavity ($\mathrm{Re}=2500$}), and NACA0012 ($\mathrm{Re}=100$, $\alpha=15^\circ$).
\label{fig:dt_compare}
\end{figure*}

The influence of the outer time-step size $\Delta t$ on FDTO convergence is also investigated. Fig.~\ref{fig:dt_compare} reports the relative $L_2$ error histories for $\Delta t\in\{0.01,\,0.05,\,0.1\}$ under the same optimization configuration.

For the lid-driven cavity at $\mathrm{Re}=2500$, increasing $\Delta t$ yields faster and deeper error reduction. With $\Delta t=0.1$, the error decays nearly monotonically, reaches $10^{-2}$ within a few hundred epochs, and then remains on a low plateau. By comparison, $\Delta t=0.05$ converges more slowly and levels off at a higher error, while $\Delta t=0.01$ shows only marginal improvement over 600 epochs and stays above $10^{-1}$. These trends suggest that an overly small $\Delta t$ leads to under-updated outer steps and consequently slow practical convergence for this case.

For the NACA0012 case ($\mathrm{Re}=100$, $\alpha=15^\circ$), the trend is reversed.
While $\Delta t=0.01$ and $\Delta t=0.05$ both yield stable convergence to a low-error regime (around $10^{-2}$) after the rapid initial drop in the error, the large step size $\Delta t=0.1$ becomes unstable: despite an early decrease, the error gradually increases and eventually grows to a significantly higher level.
This indicates that, an overly aggressive outer update can overshoot the stable descent region and amplify discretization/coupling errors in the optimization-driven update.

Fig.~\ref{fig:dt_compare} reveals a robustness-speed trade-off in selecting $\Delta t$. Larger $\Delta t$ can accelerate convergence when the dynamics are relatively well behaved (e.g., the lid-driven cavity), but it may reduce stability for more challenging cases (e.g., airfoil flow). In practice, a moderate or smaller $\Delta t$ is often preferable for robustness, and an adaptive schedule that decreases $\Delta t$ over the optimization can balance early-stage speed with late-stage stability.
\subsubsection{Optimizer selection}
Three optimizers (Adam, L-BFGS, and SOAP) are compared for minimizing the discrete loss under identical discretization, optimization settings, and stopping criteria. For each case, the relative $L_2$ error of the predicted solution is tracked against the corresponding reference (analytical/CFD) solution as optimization proceeds.
\begin{figure*}[!ht]
\centering
\includegraphics[width=0.96\textwidth]{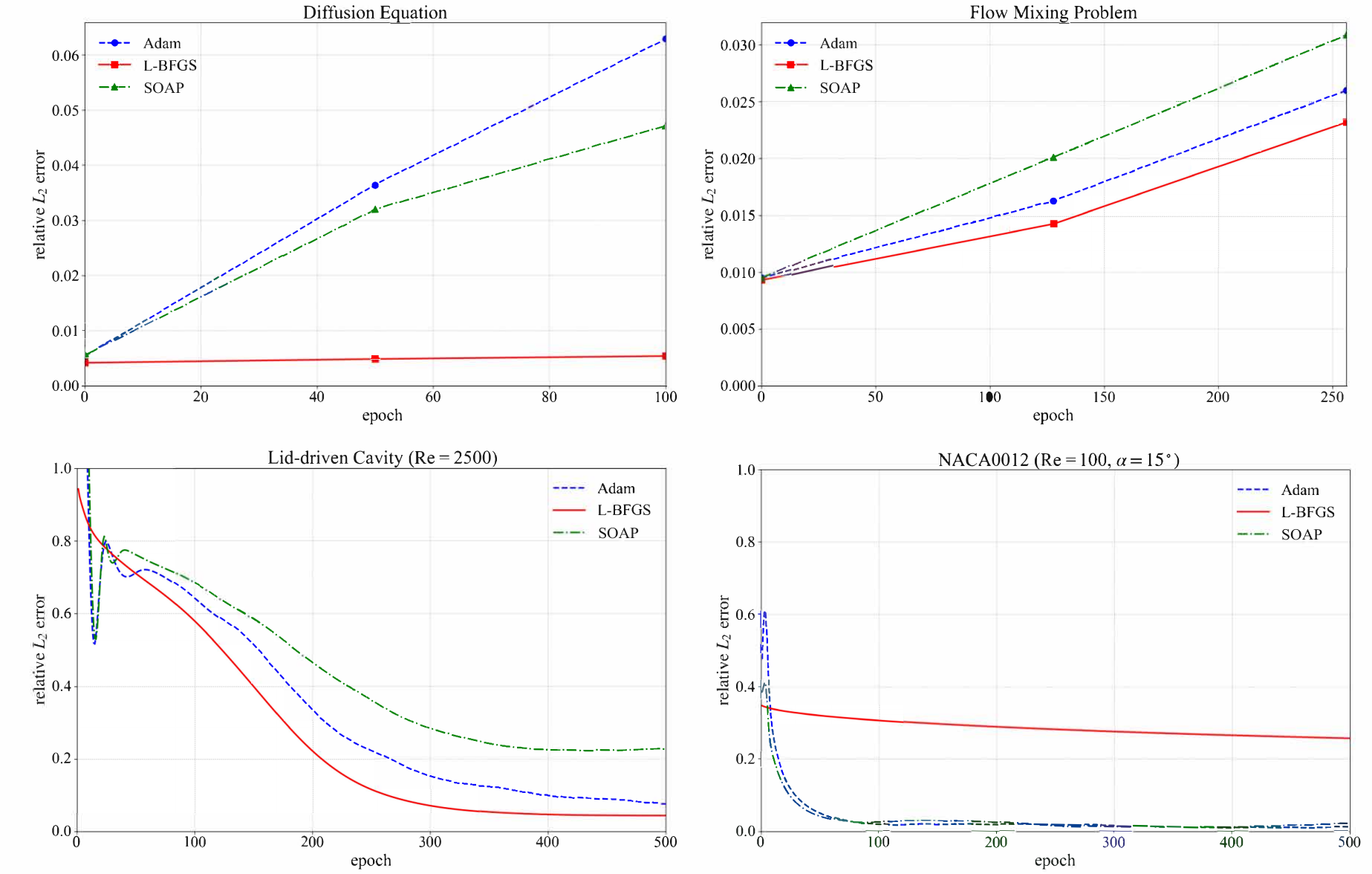}
\caption{Effect of optimizer choice (Adam, L-BFGS, and SOAP) across four cases: diffusion equation, flow-mixing problem, lid-driven cavity ($\mathrm{Re}=2500$}), and NACA0012 ($\mathrm{Re}=100$, $\alpha=15^\circ$).
\label{fig:optimizer_compare}
\end{figure*}

Fig.~\ref{fig:optimizer_compare} shows that optimizer behavior is strongly problem dependent. For the diffusion and flow-mixing problem, L-BFGS consistently maintains the lowest error throughout training, indicating that quasi-Newton updates are effective for these smooth objectives and can better control error growth over epochs. For the lid-driven cavity case, all optimizers reduce the error, but L-BFGS converges faster and reaches a lower final plateau, while Adam converges more slowly and SOAP exhibits a higher-error saturation, suggesting reduced effectiveness when the optimization becomes increasingly stiff in later epochs. In contrast, for the NACA0012 case, Adam and SOAP rapidly drive the error down to a low level within the early epochs, whereas L-BFGS stalls at a substantially higher error. 

No single optimizer dominates across all regimes: L-BFGS is highly competitive on smooth PDE cases and the lid-driven cavity flow, while adaptive first-order methods (Adam) and preconditioned variants (SOAP) are more robust for the airfoil case.
\subsubsection{Grid-Resolution Study}
\begin{figure}[!ht]
\centering
\includegraphics[width=0.48\textwidth]{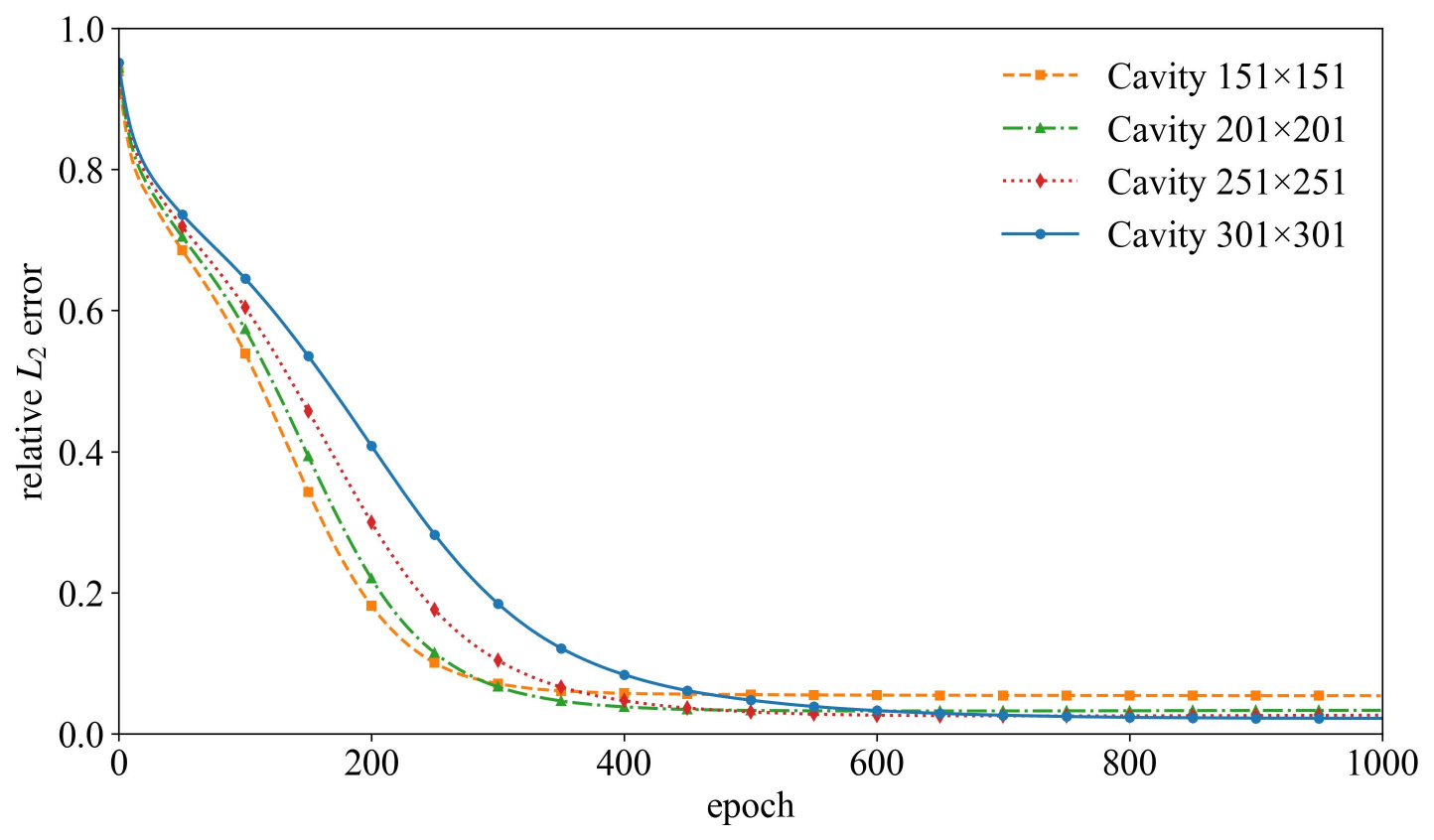}
\caption{Grid-resolution study for lid-driven cavity flow at $\mathrm{Re}=2500$: relative $L_2$ error versus epoch on $151\times151$, $201\times201$, $251\times251$, and $301\times301$ grids.}
\label{fig:loss_comparison}
\end{figure}

A grid-resolution study is conducted to assess the sensitivity of FDTO to spatial discretization and the robustness of its convergence across grid sizes. Fig.~\ref{fig:loss_comparison} reports the relative $L_2$ error histories for the lid-driven cavity at $\mathrm{Re}=2500$ on $151\times151$, $201\times201$, $251\times251$, and $301\times301$ grids. Across all resolutions, the error decreases smoothly and transitions into a stable plateau, indicating robust convergence with respect to grid refinement. The convergence behavior, however, exhibits a clear accuracy-effort trade-off: coarser grids reduce the error more rapidly in the early stage but settle at higher plateau levels, whereas finer grids require more epochs to reach the turning point yet achieve lower final errors. These trends suggest that higher resolutions expose richer small-scale features and stricter residual consistency, which increases optimization difficulty; stronger optimization settings (e.g., larger inner iterations budgets or adaptive schedules) can therefore be beneficial to fully realize the accuracy gains on refined grids.
\subsubsection{Time-Stepping Ablation}
\begin{figure}[!ht]
\centering
\includegraphics[width=0.48\textwidth]{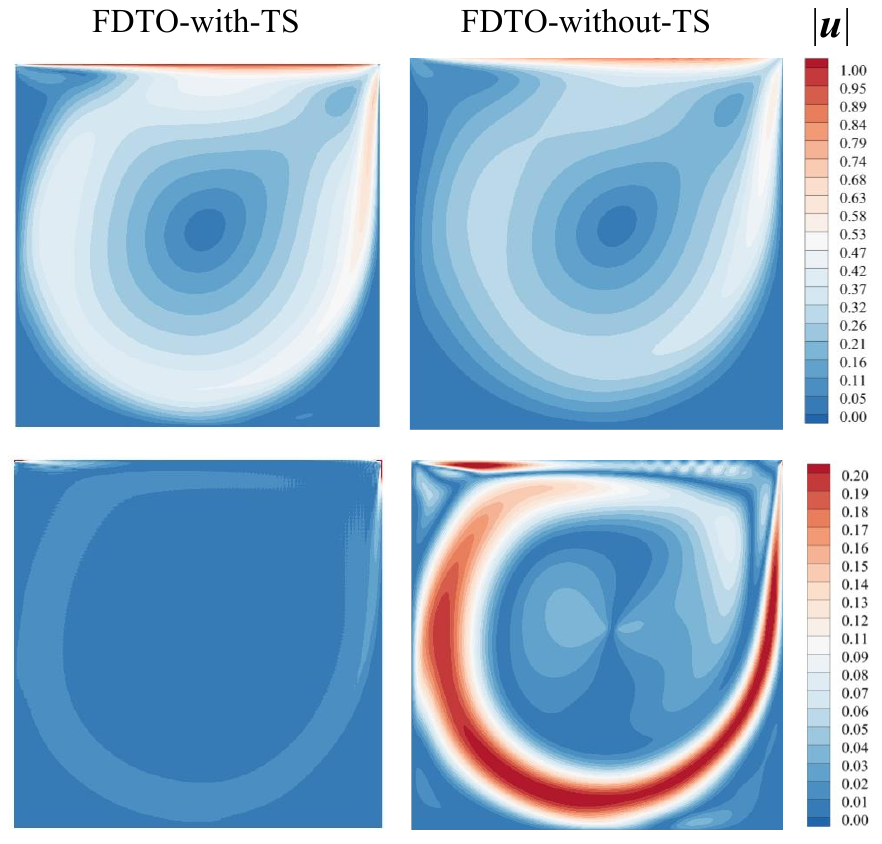}
\caption{Effect of TS optimization on lid-driven cavity flow at $\mathrm{Re}=2500$ on a $201\times201$ grid: velocity-magnitude contours (top) and absolute error maps (bottom) for FDTO with and without TS.}
\label{fig:TS_effect}
\end{figure}

\begin{figure}[!ht]
\centering
\includegraphics[width=0.48\textwidth]{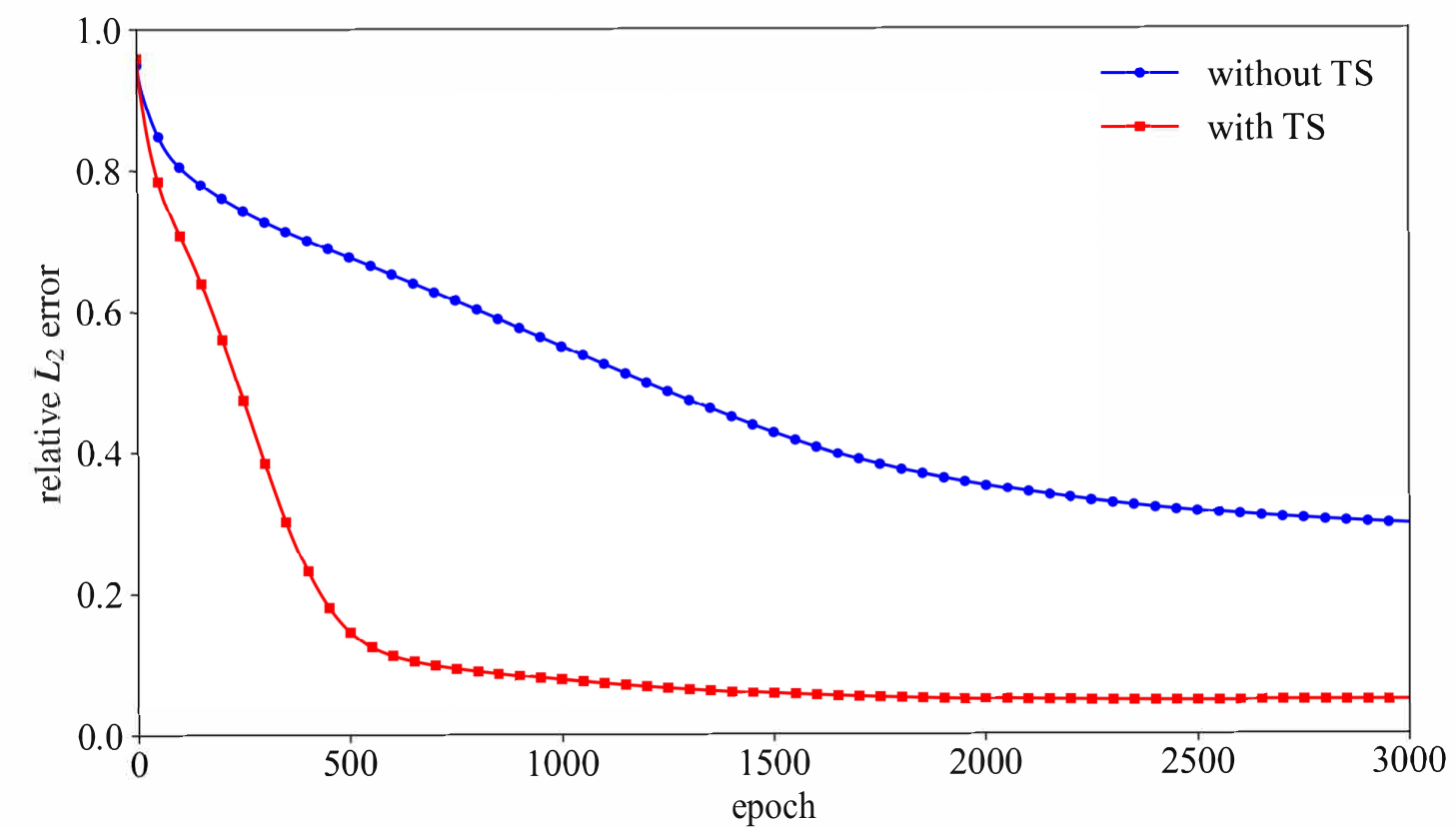}
\caption{Effect of time-stepping optimization on lid-driven cavity flow at $\mathrm{Re}=2500$ on a $201\times201$ grid: convergence of relative $L_2$ error with and without TS.}
\label{fig:cavity_ablation}
\end{figure}

The impact of the proposed time-stepping (TS) optimization on stability and temporal coherence is assessed within the optimization-driven time-marching procedure. As shown in Fig.~\ref{fig:TS_effect} for the lid-driven cavity at $\mathrm{Re}=2500$ on a $201\times201$ grid, the velocity-magnitude contours (top row) recover the dominant recirculation pattern in both settings; the absolute error maps (bottom row), however, reveal pronounced differences. With TS enabled, errors remain uniformly small and spatially smooth, whereas disabling TS leads to substantially larger and more structured discrepancies, including a distinct high-error band around the primary vortex and amplified errors near wall-adjacent shear layers. This advantage is further quantified by the convergence histories: Fig.~\ref{fig:cavity_ablation} shows that the time-marching formulation converges faster and attains lower errors than the time-independent variant. These results indicate that incorporating temporal dynamics improves optimization stability and helps resolve nonlinear flow structures. TS therefore serves as an effective stabilization mechanism, enforcing temporally consistent evolution, reducing oscillatory updates during time marching, and improving robustness and accuracy in convection-dominated, high-Reynolds-number regimes.
\subsubsection{Ablation of N-C-N Stabilization}
\begin{figure*}[!ht]
\centering
\includegraphics[width=0.6\textwidth]{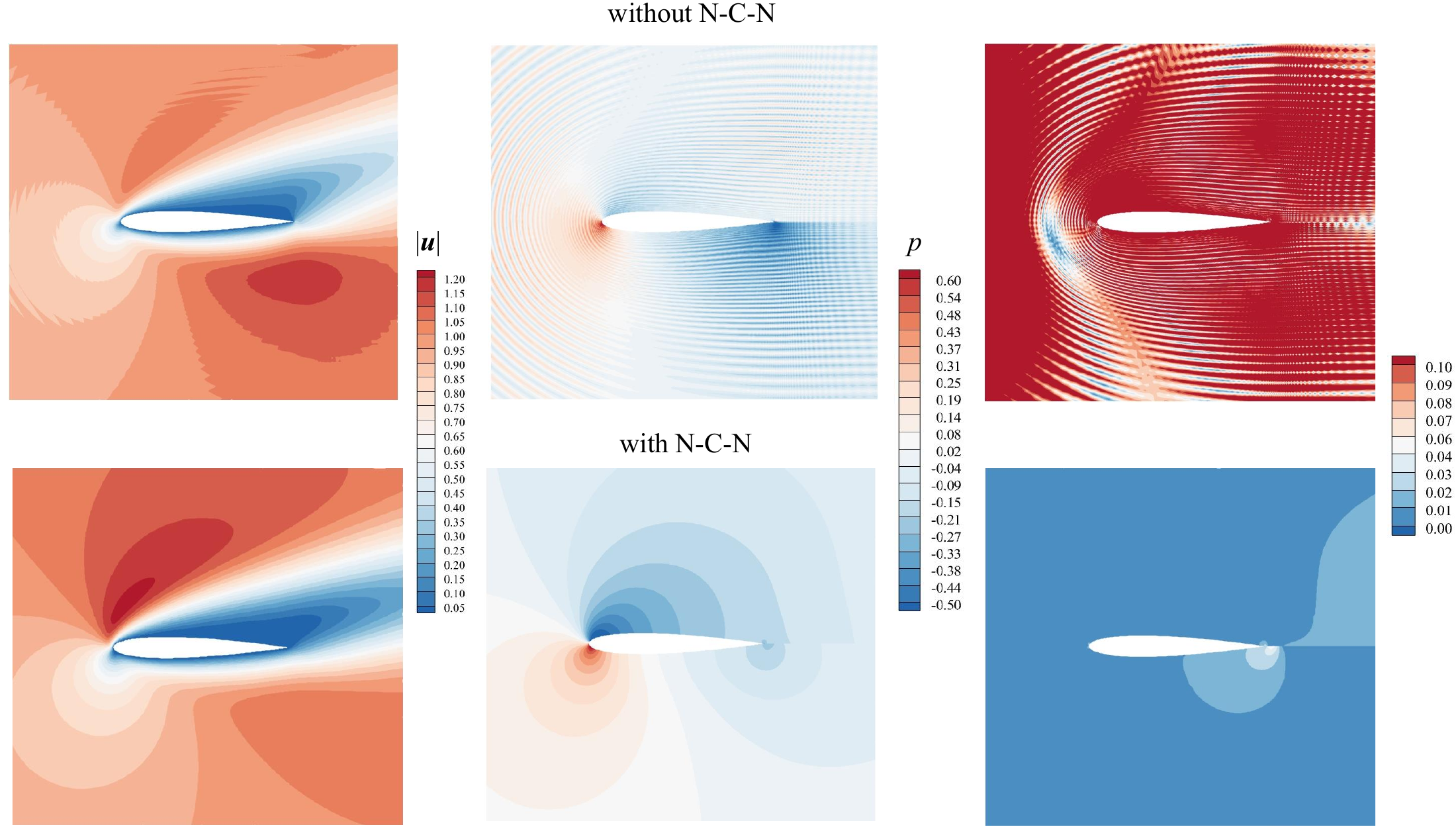}
\caption{Ablation study of the N-C-N treatment for an airfoil case: comparison of velocity magnitude (left), pressure (middle), and absolute pressure error (right) without (top row) and with (bottom row) N-C-N.}
\label{fig:NCN}
\end{figure*}

The effect of the N-C-N averaging treatment on suppressing spurious oscillations in convection-dominated regimes is assessed next. Fig.~\ref{fig:NCN} shows that, without N-C-N averaging, high-frequency oscillations accumulate during optimization-driven time marching, producing irregular pressure contours and enlarged pressure-error regions near the trailing edge and within the wake. When the N-C-N operator is applied to the discrete state $(\bm{u},p)$, field regularity improves markedly and pressure errors are substantially reduced across the domain. These results indicate that the lightweight, fully discrete N-C-N stabilization improves optimization stability and strengthens pressure-reconstruction robustness in wake-dominated flows, consistent with the design rationale discussed in the Introduction.
\subsubsection{Study on convection term discretization scheme}

Since FDTO optimizes the residuals of a discrete PDE system, its performance may be affected by the numerical discretization used to construct the residuals. To examine the influence of convection term discretization, we conducted an additional ablation study for the lid-driven cavity flow at $\mathrm{Re}=2500$ on a $201\times201$ structured grid. All cases were performed using the same settings, including the L-BFGS optimizer, a learning rate of 1.0, and $\Delta t = 0.1$. Three convection term discretization schemes were compared: the second-order central difference scheme, the fourth-order central difference scheme, and the QUICK scheme.

\begin{figure}[!ht]
\centering
\includegraphics[width=0.48\textwidth]{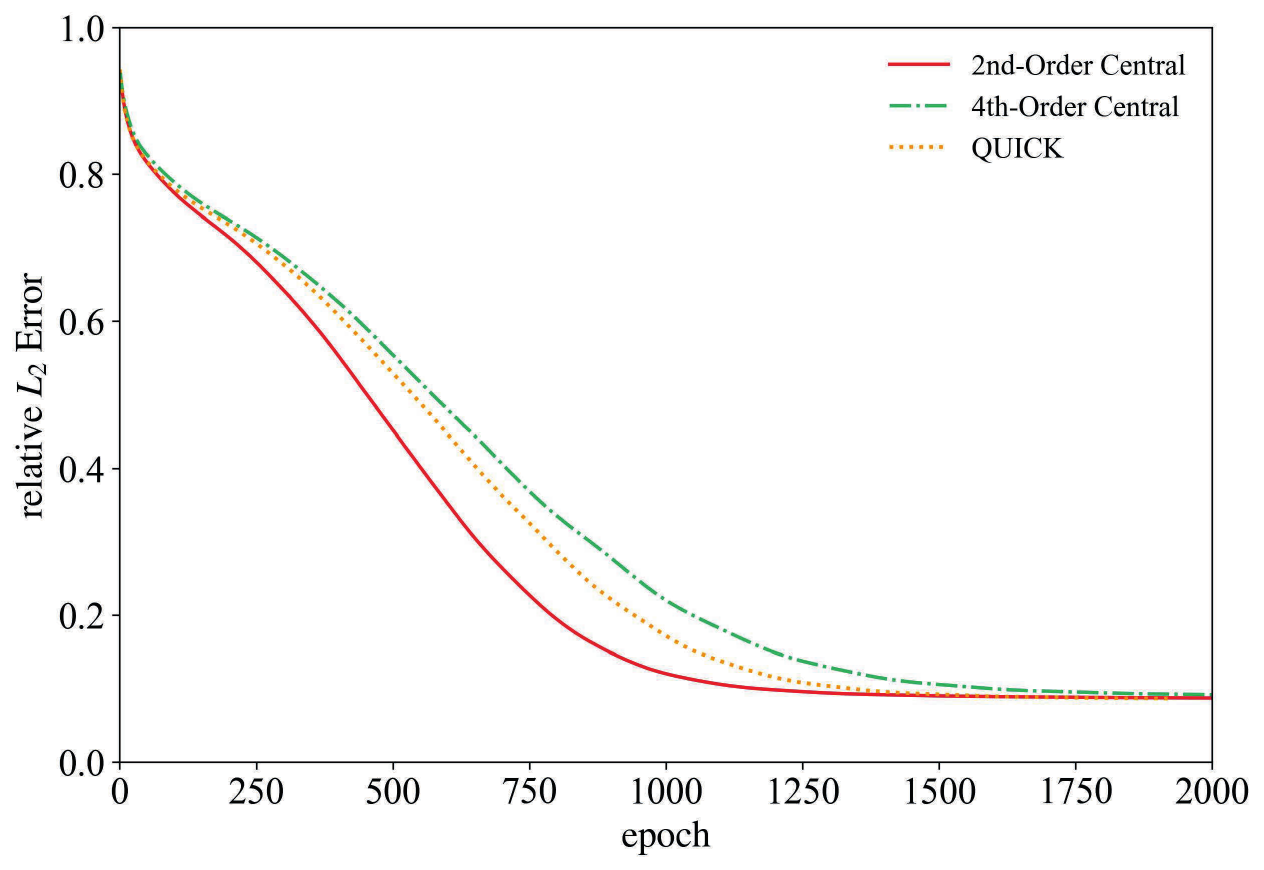}
\caption{Relative $L_2$ error histories obtained using different convection term discretization schemes: second-order central, fourth-order central, and QUICK schemes.}
\label{fig:fdto_scheme_error}
\end{figure}

As shown in Fig.~\ref{fig:fdto_scheme_error}, all three schemes lead to a clear decrease in the relative \(L_2\) error during optimization. The second-order central scheme exhibits faster error decay in the early and intermediate stages, while the QUICK and fourth-order central schemes also converge stably and eventually reach comparable error levels. These results indicate that the proposed FDTO framework is not restricted to a specific finite-difference stencil. Although the choice of convection term discretization can influence the convergence rate, the overall optimization behavior remains robust across different schemes.
\begin{table}[!ht]
\centering
\caption{Average per-iteration time costs of second-order central, fourth-order central, and QUICK discretization schemes.}
\begin{tabular}{cc}
\hline
\multicolumn{1}{l}{Discretization Strategies} & \multicolumn{1}{l}{Time costs per iteration (s)} \\ \hline
2nd-Order Central                             & 0.84                                             \\
4th-Order Central                             & 0.87                                             \\
QUICK                                         & 0.99                                             \\ \hline
\end{tabular}
\label{tab:scheme_time_cost}
\end{table}

To further assess the computational overhead introduced by different convection term discretization schemes, the average time cost per iteration was recorded, as summarized in Tab.~\ref{tab:scheme_time_cost}. The second-order central scheme requires the lowest computational cost, with an average time of 0.84 s per iteration. The fourth-order central scheme slightly increases the cost to 0.87 s per iteration, mainly due to the use of a wider stencil. The QUICK scheme shows the highest cost, reaching 0.99 s per iteration, because it involves an upwind-biased reconstruction and additional stencil operations. Nevertheless, the computational costs of the three schemes remain within the same order of magnitude, indicating that higher-order or upwind-biased discretization strategies can be incorporated into the FDTO framework with acceptable additional overhead.

This observation suggests that FDTO has the potential to be further combined with higher-order, upwind-biased, or shock-capturing discretization schemes in future studies, especially for more complex flow problems involving stronger convection, sharp gradients, or discontinuities.

\section{Conclusions}
This work proposes FDTO, a finite-differentiable, discrete-optimization framework for PDE solving that optimizes the unknown fields directly over discrete degrees of freedom. FDTO combines classical discretizations with an optimization-driven time-marching scheme and incorporates a lightweight, fully discrete smoothing treatment to suppress error accumulation in convection-dominated regimes. As a result, the method improves convergence stability while retaining geometric adaptability on body-fitted grids.

Sec.~\ref{sec:res} demonstrates the effectiveness of FDTO primarily on incompressible Navier--Stokes problems, including lid-driven cavity flows across a wide Reynolds-number range, external airfoil aerodynamics, and a cylinder wake on a multi-block structured grid. Diffusion and nonlinear flow-mixing problem are also included to illustrate broader applicability. Across these settings, FDTO delivers accurate field predictions with smooth, stable convergence and substantially lower GPU memory usage than representative neural solvers. Moreover, it maintains robust pressure reconstruction in wake-dominated regions and preserves cross-block coherence without interface-aligned artifacts. Collectively, these results suggest that FDTO provides a practical route for solving convection-dominated flows and other PDEs through a purely discrete optimization formulation.

Several limitations remain. Optimization becomes more demanding on finer grids, motivating grid-consistent normalization/weighting and continuation strategies such as coarse-to-fine initialization. Extending the framework to higher-Reynolds-number regimes and fully three-dimensional geometries will likely require stronger stabilization and preconditioning, together with scalable parallel implementations. Addressing these challenges is an important direction toward large-scale, engineering applications.
\section*{Author contributions}
CRediT: Yali Luo:Investigation, Methodology, Software, Validation, Visualization, Writing-original draft; Yiye Zou: Investigation, Methodology, Software; Heng Zhang: Validation, Visualization; Mingjie Zhang: Writing review \& editing; Gang Wei: Writing review \& editing; Jingyu Wang: Funding acquisition, Resources, Supervision, Writing-review \& editing; Xiaogang Deng:Project administration, Writing-review \& editing
\section*{Disclosure statement}
No potential conflict of interest was reported by the author(s).
\section*{Funding}
This work was supported by Sichuan Science and Technology Program (Project No.2021ZDZX0001), the Open Funding of
National key laboratory of Fundamental Algorithms and Models for Engineering Simulation, and the Sichuan University Interdisciplinary Innovation Fund.
\section*{Data availability statement}
The code and grid used in this paper can be found in the GitHub repository:\url{https://github.com/My-git96/FDTO_PDE}
\section*{Declaration of generative AI and AI-assisted technologies in the manuscript preparation process}
During the preparation of this work the author(s) used ChatGPT in order to improve phrasing. After using this tool/service, the author(s) reviewed and edited the content as needed and take(s) full responsibility for the content of the publication.






\end{document}